\documentclass[12pt]{article}
\usepackage{amsfonts}
\usepackage{amsmath}
\usepackage{amssymb}
\usepackage{graphicx}
\usepackage{pst-node}
\usepackage{tikz-cd} 
\usepackage{color}
\usepackage{slashed} 
\usepackage[margin=3cm]{geometry}

\def \be {\begin{equation}}
\def \ee {\end{equation}}
\def \bea {\begin{eqnarray}}
\def \eea {\end{eqnarray}}
\def \nn {\nonumber}

\def \rr {\raise.35ex\hbox{\small $\prime$}\kern-.17em{\mbox{\large $\imath$}}}

\def \dels {\partial\kern-.6em /\kern.1em}
\def \As {{A\kern-.5em / \kern.5em}}
\def \Ds {D\kern-.7em / \kern.5em}

\def \ks {k\kern-.5em /}
\def \ls {l\kern-.5em /}

\newcommand{\hide}[1]{}

\reversemarginpar
\begin{document}

\begin{titlepage}

\begin{center}

\hfill
\vskip .2in

\textbf{\LARGE
Parity Anomaly and Duality Web
\vskip.3cm
}

\vskip .5in
{\large
Chen-Te Ma \footnote{e-mail address: yefgst@gmail.com} 
\\
\vskip 1mm
}
{\sl
Department of Physics and Center for Theoretical Sciences, \\
National Taiwan University,\\ 
Taipei 10617, Taiwan, R.O.C.
}\\
\vskip 1mm
\vspace{40pt}
\end{center}
\begin{abstract}
We review the parity anomaly and a duality web in 2+1 dimensions. An odd dimensional non-interacting Dirac fermion theory is not parity invariant at quantum level. We demonstrate the parity anomaly in a three dimensional non-interacting Dirac fermion theory and a one dimensional non-interacting Dirac fermion theory. These theories can generate non-gauge invariant Abelian Chern-Simons terms at a finite temperature through an effective action. The parity anomaly also leads us to study the duality web in 2+1 dimensions, in which the 2+1 dimensional duality web begins from the conjecture of a duality between a three dimensional Dirac fermion theory and a three dimensional interacting scalar field theory at the Wilson-Fisher fixed point. We first review the duality web for the flat background, then we discuss its extension to the spin$_c$ manifold to avoid inconsistency from the spin structure. This also leads a global effect. We discuss that the composite fermions approach of the quantum Hall system also suffers from the same issue of the global description. Finally, we use perspective of the electric-magnetic duality of the Abelian gauge theory in 3+1 dimensions to study the 2+1 dimensional duality web.
\end{abstract}

\end{titlepage}

\section{Introduction}
\label{1}
Symmetry can be seen as an operation that maps an object to itself. The symmetry can be built from a mathematical group. To build a theory from phenomena, we are interested in using a mathematical group to find a symmetry transformation because it is a convenient way and a useful method. Quantum field theory is defined by using a Fock space, which is a direct sum of all particle numbers of a Hilbert space, and the expectation value of an $n$-point function. To study a geometric quantity in quantum field theory and give a convenient exploring in quantum field theory, we are interested in building an action from a symmetry. 

When an action of a theory is invariant under a symmetry, but the symmetry is not valid at quantum level or from loop correction or a regularization, then this phenomenon occurs in a theory, we say that the theory has anomaly. The parity anomaly is that an action is invariant under a parity transformation, but quantum correction breaks the parity invariance of the theory. 

The parity anomaly was found in an odd dimensional non-interacting Dirac fermion theory with the Abelian background gauge field for the flat background and zero temperature \cite{Niemi:1983rq}. The parity symmetry violating term is generated from a non-gauge invariant Abelian Chern-Simons term by computing the one-loop effective action \cite{Niemi:1983rq}. The study is also extended to the SU($N$) non-Abelian background gauge field and also shows the parity anomaly through an SU($N$) non-Abelian Chern-Simons term \cite{Redlich:1983dv}. Because the Chern-Simons terms were found by introducing a regularized fermion mass term \cite{Redlich:1983dv}, the regularized fermion mass term causes the ambiguity between the mass and regulators in their computation. However, one can use $\zeta$-regularization without introducing the regularized fermion mass to distinguish the effect of the mass and the regulators from the parity anomaly in the 2+1 dimensional Dirac fermion theory with an Abelian background gauge field \cite{GamboaSaravi:1994aq}. Hence, a regularization in odd dimensional spacetime should be the true reason for the parity anomaly \cite{GamboaSaravi:1994aq}. Parity anomaly conserving terms and parity anomaly violating terms were also systematically analyzed in the effective action from an any dimensional non-interacting Dirac fermion theory under the large mass and the small mass expansions \cite{Deser:1997gp}. 

By introducing a finite temperature, the one-loop effective action of a 2+1 dimensional Dirac fermion theory with an Abelian background gauge field generates a non-gauge invariant Abelian Chern-Simons term with a temperature dependent term \cite{Deser:1997nv}. The temperature dependent term lets the coefficient of the Abelian Chern-Simons term being a non-discrete value. One showed that the issue can be solved in the 2+1 dimensional non-interacting Dirac fermion with the Abelian background gauge field at any temperature \cite{Deser:1997nv}. They also computed the exact effective action in a 0+1 dimensional non-interacting Dirac fermion theory with the Abelian background gauge field \cite{Deser:1997nv} to demonstrate that the issue comes from the perturbation. This issue of the perturbation was also demonstrated from the exact effect action in a 2+1 dimensional non-interacting Dirac fermion theory for the case of the Abelian background gauge field \cite{Fosco:1997ei} and the SU($N$) non-Abelian gauge field \cite{Fosco:1997vu} when one component of the gauge field only depends on time and other components of the gauge field do not depend on time. When the Dirac fermion theory coupled to the gravitation field at zero temperature, one showed the gravitational parity anomaly in the theory \cite{AlvarezGaume:1984nf}. One computed the effective action of a 2+1 dimensional non-interacting Dirac fermion theory coupled to the gravitation field to obtain that the gravitational parity anomaly is related to a gravitation Chern-Simons term \cite{Kurkov:2018pjw}. The parity anomaly was also studied on the non-orientable manifold \cite{Witten:2016cio}. 

The 2+1 dimensional Chern-Simons theory is topological quantum field theory. This theory can be quantized from the canonical quantization or the holomorphic quantization when the level is an integer \cite{Elitzur:1989nr}. Furthermore, the 2+1 dimensional Chern-Simons theory has interesting topological invariant quantities to do application in condensed matter systems \cite{Witten:1988hf}. The integer quantum Hall effect is a quantum phenomenon for the quantized Hall conductance or an integer filling factor in two dimensional electron systems. The factional quantum Hall effect is a physical phenomenon when the Hall conductance is a fractional value or the filling factor is a fractional value \cite{Laughlin:1983fy}. This was found when the filling factor is $1/(2k+1)$, where $k$ is a non-negative integer \cite{Laughlin:1983fy}. More rational filling factors could be constructed by adding quasi-particle or holes \cite{Halperin:1984fn}. The fractional quantum Hall effect could be understood as in the integral quantum Hall effect through the composite fermions \cite{Zhang:1988wy}. The 2+1 dimensional Chern-Simons term plays an important role in the formulation of the composite fermions \cite{Zhang:1988wy}. 

The 2+1 dimensional Chern-Simons theory only has gauge symmetry when the level is an integer. When one discusses the parity anomaly in a 2+1 dimensional non-interacting Dirac fermion field theory through the one-loop effective potential at zero temperature, the Chern-Simons term has one-half coefficient. Thus, this result provides that the 2+1 dimensional Dirac fermion field theory should not be gauge invariant at quantum level when one does regularization \cite{Atiyah:1975jf}. This also leads us to study a 2+1 dimensional duality web. The duality web on a lattice is also proposed with manifest gauge symmetry and statistical transmutation, which is the same as the Chern-Simons term at the continuum limit \cite{Chen:2017lkr}. The lattice fermion theory in this duality web is defined from the Wilson-Dirac fermions because the Dirac fermion theory does not have chiral symmetry in odd dimensions \cite{Chen:2017lkr}.

When two theories have a duality, it implies that two theories are equivalent through a non-trivial mapping. Thus, building a duality web can unify different theories through dualities. The duality web was first constructed in string theory and the ten dimensional supergravity theories \cite{Giveon:1994fu}. String theory and the ten dimensional supergravity theories were unified via the T-duality and the S-duality. The T-duality is a target space duality by showing equivalence between a theory with the radius $R$ of a circle and a theory with the inverse radius $1/R$ of a circle. The S-duality shows equivalence by exchanging a weak coupling constant and a strong coupling constant. The famous example of the S-duality is the electric-magnetic duality in the 3+1 dimensional Abelian gauge theory \cite{Witten:1995gf}. The procedure of the electric-magnetic duality can be extended to the $2p+2$ dimensional non-interacting $p$-form gauge field theory \cite{Ho:2015mfa}. 

The above duality web should be built at a very high energy scale. Now we introduce a 2+1 dimensional duality web at a low-energy scale. We first introduce a particle-vortex duality for the boson fields in 2+1 dimensions at the Wilson-Fisher fixed point or the infrared (IR) limit \cite{Peskin:1977kp}. The particle-vortex duality is a duality that maps fundamental particles to vortices or monopole operators at the IR limit. The simplest example is a particle-vortex duality between a fundamental field in a complex scalar field theory and a monopole operator in the Abelian-Higgs model at the IR limit \cite{Peskin:1977kp}. Then the particle-vortex duality for 2+1 dimensional Dirac fermion fields was also found in a non-interacting Dirac fermion theory and a interacting Dirac fermion theory at an IR limit \cite{Metlitski:2015eka}. The Dirac fermion field in the non-interacting Dirac fermion theory is mapped to the monopole operator in the interacting Dirac fermion theory at the IR limit \cite{Metlitski:2015eka}. The 2+1 dimensional duality web was proposed to include the above dualities from the conjecture that a non-interacting Dirac fermion theory is equivalent to an interacting scalar field theory in 2+1 dimensions at the Wilson-Fisher fixed point \cite{Karch:2016sxi}. This 2+1 dimensional duality web also showed that one dualization has inequivalent dependence of the spin structure on a spin manifold \cite{Karch:2016sxi}. The inconsistent duality between one purely boson theory and a Dirac fermion field theory coupled to a dynamical gauge field \cite{Seiberg:2016gmd}. We know that the purely boson theory can be formulated on a non-spin manifold, but the Dirac fermion theory can only be putted on a spin manifold \cite{Witten:2015aba}. Hence, the duality web should be careful about a choice of the background. To obtain a more generic duality web, one considers a spin$_c$ manifold \cite{Seiberg:2016gmd} to obtain a globally defined theory rather than considering a spin manifold based on the Atiyah-Patodi-Singer index theorem \cite{Atiyah:1975jf, Seiberg:2016rsg}. Now we also know that dualities of the 2+1 dimensional duality web can be connected to the T-operation and the S-operation of the 3+1 dimensional Abelian gauge theory \cite{Seiberg:2016gmd, Witten:2003ya}. The relation between the dualities in four dimensions and the dualities in three dimensions should let the 2+1 dimensional duality web be more reliable without using the conjecture of the 2+1 dimensional duality web \cite{Seiberg:2016gmd, Murugan:2016zal}. 

We also have other generalized dualities, which are motivated by the 2+1 dimensional duality web. For example, a different conjecture of the 2+1 dimensional duality web was proposed by promoting background fields to dynamical fields \cite{Karch:2016aux, Ma:2016yas, Ma:2017mpb}, the extension of the finite temperature in 2+1 dimensions \cite{Ma:2016yas, Ma:2017mpb}, the extension of the SO gauge group \cite{Metlitski:2016dht, Aharony:2016jvv} and the USp gauge group in the 2+1 dimensional duality web \cite{Aharony:2016jvv}, and discussion of a higher dimensional duality web from the electric magnetic duality \cite{Ma:2017mpb}. The duality web is also related to the mirror symmetry, which is a duality between supersymmetric theories. This direction leads to supersymmetric defect \cite{Hook:2013yda}, a half-filled Landau level \cite{Kachru:2015rma}, bosonization \cite{Kachru:2016rui} and other non-supersymmetric dualities \cite{Kachru:2016aon}.

\subsection{Outline}
The outline of the review is as follows. We introduce the parity anomaly from the time reversal symmetry and the Atiyah-Patodi-Singer theorem in arbitrary dimensions. Then we compute a 2+1 dimensional non-interacting Dirac fermion field theory at a finite temperature through the one-loop effective action in Sec.~\ref{2}. In additional, we also introduce an exact way to show the parity anomaly in a 0+1 dimensional non-interacting Dirac fermion theory at a finite temperature in this section.

We review the particle-vortex duality of the boson fields and also show the 2+1 dimensional duality web when we consider the flat background in Sec.~\ref{3}.  

We demonstrate that the problem of the spin structure in the 2+1 dimensional duality web and show how to build the 2+1 dimensional duality web on the spin$_c$ manifold in Sec.~\ref{4}. For convenience of readers, we also review the mathematical definition for the spin manifold and the spin$_c$ manifold. 

We review the mean-field theory and introduce quasi-particle to composite fermions approach of the quantum Hall system and also show that it has the similar issue of a global description as in the 2+1 dimensional duality web in Sec.~\ref{5}.

We review an electric-magnetic duality in the four dimensional Abelian gauge theory and also discuss a conjecture of the 2+1 dimensional duality web from the four dimensional electric-magnetic duality in Sec.~\ref{6}. 

Finally, we discuss future directions in Sec.~\ref{7}. 

\section{Parity Anomaly and Dirac Fermion Theory}
\label{2}
We first introduce the parity anomaly from the time reversal symmetry in the non-interacting Dirac fermion theory. Then we compute the one-loop effective potential to obtain the non-gauge invariant Abelian Chern-Simons term in a 2+1 dimensional non-interacting Dirac fermion theory \cite{Niemi:1983rq, Deser:1997nv}. We also introduce an exact way to understand the parity anomaly in a one dimensional non-interacting Dirac fermion theory \cite{Deser:1997nv}.

\subsection{Time Reversal Symmetry and Dirac Fermion Theory}
We start from the action of the massless Dirac fermion theory on the $D-1$ dimensional manifold
\bea
S_{\mathrm{mldf}}=\int_W d^{D-1}x\ \bar{\psi}\gamma^{\mu}(i\partial_{\mu}-A_{\mu})\psi,
\eea
where $\gamma^{\mu}$ is the Gamma matrix:
\bea
\gamma^0\equiv\begin{pmatrix} 1 & 0
 \\ 0 & -1 \end{pmatrix}, 
 \qquad 
 \gamma^1\equiv\begin{pmatrix} 0 & i
 \\ i & 0 \end{pmatrix}, 
 \qquad
 \gamma^2\equiv\begin{pmatrix} 0 & 1
 \\ -1 & 0 \end{pmatrix}, 
\eea
$A_{\mu}$ is a U(1) non-dynamical gauge field, $\bar{\psi}\equiv\psi^{\dagger}\gamma_0$, $D$ is a number of dimensions of a worldvolume. Note that the spacetime indices are labeled by the Greek indices ($\mu, \nu, \cdots$). The manifold $X$ is a $D$ dimensional manifold and $W$ is its $D-1$ dimensional boundary manifold and also these manifolds are the Riemannian manifolds with a spin structure. The Gamma matrix $\gamma$ satisfies:
\bea
\{\gamma_{\mu}, \gamma_{\nu}\}=2\eta_{\mu\nu}, \qquad \eta_{\mu\nu}\equiv\mbox{diag}(1, -1, -1,\cdots).
\eea
The time reversal transformation $T$ is defined by:
\bea
T\big(A_0(t, \vec{x})\big)\equiv A_0(-t, \vec{x}), \qquad T\big( A_i(t, \vec{x})\big)\equiv -A_i(-t, \vec{x}),
\eea
in which the spatial indices are labeled by $a-z$. The charge conjugation $C$ is defined by:
\bea
C\big(A_0(t, \vec{x})\big)\equiv A_0(t, \vec{x}), \qquad C\big( A_i(t, \vec{x})\big)\equiv -A_i(t, \vec{x}).
\eea
Thus, we obtain:
\bea
CT\big(A_0(t, \vec{x})\big)\equiv A_0(-t, \vec{x}), \qquad CT\big( A_i(t, \vec{x})\big)\equiv A_i(-t, \vec{x}).
\eea
To consider the discrete symmetries on the Dirac fermion theory conveniently, we would like to write the Dirac fermion field $\psi$ in terms of the Majorana fermion fields ($\lambda_1$ and $\lambda_2$)
\bea
\psi\equiv\frac{(\lambda_1+i\lambda_2)}{\sqrt{2}}.
\eea
The time reversal transformation acting on the Majorana fermion fields is defined as:
\bea
T\big(\lambda_1(t, \vec{x})\big)\equiv\gamma_2\lambda_1(-t, \vec{x}), \qquad T\big(\lambda_0(t, \vec{x})\big)\equiv-\gamma_2\lambda_0(-t, \vec{x}).
\eea
Therefore, we can deduce
\bea
T\big(\psi(t, \vec{x})\big)&=&\gamma_2\psi(-t, \vec{x}\big).
\eea
Then the charge conjugation acting on the fermion fields is defined as:
\bea
C\big(\lambda_1(t, \vec{x})\big)=\lambda_1, \qquad C\big(\lambda_2(t, \vec{x})\big)=-\lambda_2(t, \vec{x}), \qquad C\big(\psi(t, \vec{x}\big)=\psi^{\dagger}.
\eea
Hence, we get:
\bea
CT\big(\lambda_1(t, \vec{x})\big)&=&\gamma_2\lambda_1(-t, \vec{x}), \qquad CT\big(\lambda_2(t, \vec{x})\big)=\gamma_2\lambda_2(-t, \vec{x}), 
\nn\\
 CT\big(\psi(t, \vec{x})\big)&=&\gamma_2\psi^{\dagger}(-t, \vec{x}).
\eea
The partition function of the massless Dirac fermion theory is $\det\big(i\gamma^{\mu}(\partial_{\mu}+A_{\mu})\big)$. Because the matrix $i\gamma^{\mu}(\partial_{\mu}+A_{\mu})$ only has real eigenvalues, the partition function should be real. We need to do regularization because the matrix has infinite negative eigenvalues. Thus, defining a sign of the partition function should suffer from the issue. Precisely, the issue is how we compute the determinant of the matrix
\bea
\det\big(i\gamma^{\mu}(\partial_{\mu}-A_{\mu})\big)=(-1)^{n_-}\bigg|\det\big(i\gamma^{\mu}(\partial_{\mu}-A_{\mu})\big)\bigg|,
\eea
where $n_-$ is a number of negative eigenvalues of the matrix $i\gamma^{\mu}(\partial_{\mu}-A_{\mu})$. We also define that $n_+$ is a number of positive eigenvalues of the matrix $i\gamma^{\mu}(\partial_{\mu}-A_{\mu})$ and $N=n_++n_-$ is the total number of eigenvalues of the matrix $i\gamma^{\mu}(\partial_{\mu}-A_{\mu})$. When we do regularization, we replace a number of negative engenvalues  $n_-$ by $n_--N/2=(n_--n_+)/2$. The subtraction still gives an ill-defined result, but a consistent regularization should give a unique outcome. We introduce the regulator
\bea
\eta=\lim_{\epsilon\rightarrow 0}\sum_i\exp(-\epsilon|\lambda_i|)\mbox{sign}(\lambda_i),
\eea
where
\bea
\mbox{sign}(\lambda)= \left\{\begin{array}{ll}
                 1, & \mbox{if $\lambda\ge 0$}, \\  
                 -1, & \mbox{if $\lambda<0$},
                \end{array} \right.
\eea
to do regularization \cite{Seiberg:2016rsg}. Therefore, we can regularize $(n_--n_+)/2$ by replacing it with $-\eta/2$ and also replace $(-1)^{n_-}=\exp(\pm i\pi n_-)$ by $\exp(\mp i\pi\eta/2)$. The regularized partition function of the massless Dirac fermion theory is given by
\bea
Z_{\mathrm{rmd}}=|\det\big(i\gamma^{\mu}(\partial_{\mu}-A_{\mu})\big)|\exp(\mp i\pi\eta/2).
\eea
From the regularization, the sign of the exponent comes from the sign of the regulator. Indeed, the non-real partition function should reflect that the regularization breaks the time reversal symmetry \cite{Atiyah:1975jf, Seiberg:2016rsg}. To preserve the time reversal symmetry at quantum level, we consider two Dirac fermion fields with the opposite sign of the regulator. This leads a real and positive regularized partition function. 

We also use the Atiyah-Patodi-Singer theorem \cite{Atiyah:1975jf}
\bea
\frac{1}{768\pi^2}\int_X d^4x\ \epsilon^{\mu\nu\rho\sigma}R_{\mu\nu}R_{\rho\sigma}+\frac{1}{32\pi^2}\int_X d^4x\ \epsilon^{\mu\nu\rho\sigma}F_{\mu\nu}F_{\rho\sigma}-\frac{\eta}{2}=J,
\eea
where
\bea
R_{\mu\nu}&\equiv&\partial_{\delta}\Gamma^{\delta}_{\nu\mu}-\partial_{\nu}\Gamma^{\delta}_{\delta\mu}
+\Gamma^{\delta}_{\delta\lambda}\Gamma^{\lambda}_{\nu\mu}
-\Gamma^{\delta}_{\nu\lambda}\Gamma^{\lambda}_{\delta\mu}, \qquad
 \Gamma^{\mu}_{\nu\delta}\equiv\frac{1}{2}g^{\mu\lambda}\bigg(\partial_{\delta}g_{\lambda\nu}+\partial_{\nu}g_{\lambda\delta}
-\partial_{\lambda}g_{\nu\delta}\bigg),
\nn\\
\eea
$J$ is an integer and $X$ is a four dimensional manifold, to discuss the time reversal symmetry when $D=4$.

Thus, we obtain
\bea
2\mbox{CS}_{\mathrm{grav}}+2\pi\bigg(\frac{1}{32\pi^2}\int_X d^4x\ \epsilon^{\mu\nu\rho\sigma}F_{\mu\nu}F_{\rho\sigma}\bigg)=\pi\eta, \qquad \mbox{mod}\ 2\pi Z ,
\eea
in which the action of the gravitation Chern-Simons term is defined by
\bea
\mbox{CS}_{\mathrm{grav}}&\equiv&\frac{1}{96\pi}\int d^3x\ \epsilon^{\mu\nu\rho}\bigg(\Gamma^{\sigma}_{\mu\delta}\partial_{\nu}\Gamma^{\delta}_{\rho\sigma}+\frac{2}{3}\Gamma^{\sigma}_{\mu\delta}\Gamma^{\delta}_{\nu\gamma}\Gamma^{\gamma}_{\rho\sigma}\bigg)
\eea
and the other term is the action of the Abelian Chenr-Simons theory with the level one:
\bea
2\pi\bigg(\frac{1}{32\pi^2}\int_X d^4x\ \epsilon^{\mu\nu\rho\sigma}F_{\mu\nu}F_{\rho\sigma}\bigg)
=\frac{1}{4\pi}\int_M d^3x\ \epsilon^{\mu\nu\rho}A_{\mu}\partial_{\nu}A_{\rho}\equiv S_{\mathrm{CS}},
\eea
where the manifold $M$ is the boundary of the manifold $X$. The Atiyah-Patodi-Singer theorem in arbitrary dimensions \cite{Atiyah:1975jf} can be written as
\bea
2\mbox{CS}_{\mathrm{grav}}+S_{\mathrm{CS}}=\pi\eta, \qquad \mbox{mod}\ 2\pi Z.
\eea

When we consider odd $D$, the regulator $\eta$ should be an even integer because the gravitation Chern-Simons and the Chern-Simons terms vanish. The regularized partition function of the massless Dirac fermion theory should be real. In the case of the even $D$ case, the regularized partition function of the massless Dirac fermion theory with the time reversal symmetry can be
\bea
&&|\det\big(i\gamma^{\mu}(\partial_{\mu}-A_{\mu})\big)|\exp\bigg(\mp i\pi\frac{\eta}{2}\bigg)\exp\bigg\lbrack\pm i\bigg(S_{\mathrm{grav}}+\frac{S_{\mathrm{CS}}}{2}\bigg)\bigg\rbrack
\nn\\
&=&|\det\big(i\gamma^{\mu}(\partial_{\mu}-A_{\mu})\big)|(-1)^J
\nn\\
\eea
by including the action of the Chern-Simons theory and the action of the gravitation Chern-Simons theory.

\subsection{One-Loop Effective Action in a 2+1 Dimensional Non-Interacting Dirac Fermion Field Theory}
We consider the action of the non-interacting Dirac fermion theory in the three dimensional Minkowski spacetime
\bea
S_{\mathrm{fermionm}}=\int d^3x\ \bigg(i\bar{\psi}\gamma^{\mu}(\partial_{\mu}+iA_{\mu})\psi+m\bar{\psi}\psi\bigg).
\eea
The action of the non-interacting Dirac fermion theory in the three dimensional Euclidean spacetime is
\bea
S_{\mathrm{fermioneu}}=\int d^3x\ \bigg(\bar{\psi}(i\gamma_{\mu}\partial_{\mu}-\gamma_{\mu}A_{\mu})\psi-m\bar{\psi}\psi\bigg)
\eea
through:
\bea
t\rightarrow -it, \qquad \gamma^0\rightarrow i\gamma^0, \qquad \gamma^i\rightarrow\gamma^i, \qquad A_0\rightarrow iA_0, 
\qquad A_i\rightarrow A_i.
\eea
Thus, the one-loop effective action of the non-interacting Dirac fermion theory ($S_{\mathrm{fermioneu}}$) is
\bea
S_{\mathrm{eff}}[A; m]\sim \ln\mbox{det}\bigg(i\gamma_{\mu}(\partial_{\mu}+iA_{\mu})-m\bigg).
\eea
Through the low energy expansion, we obtain the one-loop effective action
\bea
S_{\mathrm{eff}}[A; m]&\sim&\mbox{Tr}\ln\bigg(i\gamma_{\mu}\partial_{\mu}-m\bigg)
+\mbox{Tr}\bigg(\frac{1}{i\gamma_{\mu}\partial_{\mu}-m}\gamma_{\nu}A_{\nu}\bigg)
\nn\\
&&-\frac{1}{2}\mbox{Tr}\bigg(\frac{1}{i\gamma_{\mu}\partial_{\mu}-m}\gamma_{\nu}A_{\nu}\frac{1}{i\gamma_{\rho}\partial_{\rho}-m}\gamma_{\delta}A_{\delta}\bigg)+\cdots\ .
\eea
The parity anomaly comes from the third term of the expansion of the one-loop effective action. This term can be rewritten as
\bea
&&-\frac{1}{2}\mbox{Tr}\bigg(\frac{1}{i\gamma_{\mu}\partial_{\mu}-m}\gamma_{\nu}A_{\nu}\frac{1}{i\gamma_{\rho}\partial_{\rho}-m}\gamma_{\delta}A_{\delta}\bigg)
\nn\\
&=&\frac{1}{2}\int\frac{d^3p}{(2\pi)^3}\ \bigg(A_{\mu}(p)\Gamma_{\mu\nu}(p, m)A_{\nu}(p)\bigg),
\eea
where
\bea
\Gamma_{\mu\nu}(p, m)=-\int\frac{d^3k}{(2\pi)^3}\ \mbox{Tr}\bigg(\gamma_{\mu}\frac{\gamma_{\rho}(p_{\rho}+k_{\rho})+m}{(p+k)^2+m^2}\gamma_{\nu}\frac{-\gamma_{\delta}k_{\delta}-m}{k^2+m^2}\bigg).
\eea
We used the identity
\bea
\mbox{Tr}\bigg(\gamma_{\mu}\gamma_{\nu}\gamma_{\rho}\bigg)=2\epsilon_{\mu\nu\rho}
\eea
to extract the one-loop effective action related to the parity anomaly from the term containing three Gamma matrices
\bea
\Gamma_{\mu\nu}(p, m)\longrightarrow\Gamma^{\prime}_{\mu\nu}=\epsilon_{\mu\nu\rho}\Pi(p, m)p_{\rho},
\eea
where
\bea
\Pi(p, m)=2m\int \frac{d^3k}{(2\pi)^3}\ \frac{1}{\big((p+k)^2+m^2\big)(k^2+m^2)}.
\eea
When we consider the low-energy limit $p\rightarrow 0$, we find
\bea
\Gamma^{\prime}_{\mu\nu}(p, m)\sim\frac{1}{4\pi}\frac{m}{|m|}\epsilon_{\mu\nu\rho}p_{\rho}+\cdots\ 
\eea
by computing:
\bea
\int \frac{d^3k}{(2\pi)^3}\ \frac{1}{(k^2+m^2)}&=&\int^{\infty}_0\frac{dk}{(2\pi)^2}\ \frac{k^2}{(k^2+m^2)^2}=\frac{1}{4\pi^2|m|}\int^{\frac{\pi}{2}}_{-\frac{\pi}{2}}d\theta\ \sin^2\theta
\nn\\
&=&\frac{1}{4\pi^2|m|}\int^{\frac{\pi}{2}}_{-\frac{\pi}{2}}d\theta\ \frac{1-\cos(2\theta)}{2}
=\frac{1}{8\pi|m|},
\eea
in which we used $k=|m|\tan\theta$ in the second equality. Therefore, we obtain the 2+1 dimensional non-gauge invariant Abelian Chern-Simons term from the one-loop effective action
\bea
S_{\mathrm{effeu}}^{\mathrm{CS}}=-\frac{i}{2}\frac{1}{4\pi}\frac{m}{|m|}\int d^3x\ \epsilon_{\mu\nu\rho}A_{\mu}\partial_{\nu}A_{\rho}.
\eea
The effective action in the 2+1 dimensional Minkowski spacetime is
\bea
S_{\mathrm{eff}}^{\mathrm{CS}}=\frac{1}{2}\frac{1}{4\pi}\frac{m}{|m|}\int d^3x\ \epsilon^{\mu\nu\rho}A_{\mu}\partial_{\nu}A_{\rho}.
\eea
When we introduce the finite temperature ($T$) in the one-loop effective action, we need to compute
\bea
\Pi(0, m)=2mT\sum_{n=-\infty}^{\infty}\int\frac{d^2k}{(2\pi)^2}\frac{1}{\bigg(\big((2n+1)\pi T\big)^2+\vec{k}^2+m^2\bigg)^2},
\eea
where $\vec{k}^2=k_x^2+k_y^2$, because the Dirac fermion field satisfies the anti-periodic boundary:
\bea
\psi(0, \vec{x})=-\psi(\beta, \vec{x})\equiv -\psi\bigg(\frac{1}{T}, \vec{x}\bigg),
\eea
which determines the frequency ($\omega_n$) of the Dirac fermion field:
\bea
\psi(t, \vec{x})=\sum_{n=-\infty}^{\infty}\psi(\omega_n, \vec{x})e^{i\omega_n t}, \qquad \omega_n=(2n+1)\pi T.
\eea
We first do integration for the momentum $\vec{k}$ in $\Pi(0, m)$
\bea
\Pi(0, m)=\frac{2mT}{4\pi}\sum_{n=-\infty}^{\infty}\frac{1}{\bigg(\big((2n+1)\pi T\big)^2+m^2\bigg)}.
\eea
The summation of the index $n$ needs to use the formula:
\bea
\frac{1}{\beta}\sum_{n=-\infty}^{\infty}f\big(p^0=i(2n+1)\pi T\big)&=&T\oint_{C_1}\frac{dp^0}{2\pi i}f(p^0)\frac{1}{2T}\tanh \bigg(\frac{p^0}{2T}\bigg)
\nn\\
&=&-T\oint_{C_2}\frac{dp^0}{2\pi i}f(p^0)\frac{1}{2T}\tanh \bigg(\frac{p^0}{2T}\bigg),
\nn\\
\eea
in which the closed loop $C_1$ encloses the poles $i(2n+1)\pi T$ and the other closed loop $C_2$ encloses other poles of the function $f(p^0)$. The residue of $\tanh\big(p^0/(2T)\big)$ can be determined as:
\bea
&&\lim_{p^0\rightarrow i(2n+1)\pi T}\bigg(\frac{p^0-i(2n+1)\pi T}{e^{\frac{p^0}{2T}}-e^{-\frac{p^0}{2T}}}\bigg)\bigg(e^{\frac{p^0}{2T}}-e^{-\frac{p^0}{2T}}\bigg)
\nn\\
&=&\lim_{p^0\rightarrow i(2n+1)\pi T}\frac{2T}{e^{\frac{p^0}{2T}}-e^{-\frac{p^0}{2T}}}\bigg(e^{\frac{p^0}{2T}}-e^{-\frac{p^0}{2T}}\bigg)
\nn\\
&=&2T.
\nn\\
\eea
Hence, we can obtain:
\bea
\sum_{n=-\infty}^{n=\infty}\frac{1}{\omega_n^2+m^2}&=&\sum_{-\infty}^{\infty}\frac{1}{-(p^0)^2+m^2}\bigg|_{p^0=i(2n+1)\pi T}=\oint_{C_2}\frac{dp^0}{2\pi i}\ \frac{1}{(p^0)^2-m^2}\frac{1}{2T}\tanh\bigg(\frac{p^0}{2T}\bigg)
\nn\\
&=&\frac{1}{2m}\frac{1}{2T}\tanh\bigg(\frac{m}{2T}\bigg)+\frac{1}{-2m}\frac{1}{2T}\tanh\bigg(-\frac{m}{2T}\bigg)=\frac{1}{2mT}\tanh\bigg(\frac{m}{2T}\bigg).
\nn\\
\eea
The non-gauge invariant Abelian Chern-Simons term through the one-loop effective action at the finite temperature is given by
\bea
-i\frac{1}{2}\frac{1}{4\pi}\frac{m}{|m|}\tanh\bigg(\frac{|m|}{2T}\bigg)\int_0^{\frac{1}{T}}dt\int d^2x\ \epsilon_{\mu\nu\rho}A_{\mu}\partial_{\nu}A_{\rho},
\eea 
which provides the parity anomaly at the finite temperature.
Hence, the Abelian Chern-Simons term at the finite temperature in the 2+1 dimensional Minkowski spacetime is given by
\bea
\frac{1}{2}\frac{1}{4\pi}\frac{m}{|m|}\tanh\bigg(\frac{|m|}{2T}\bigg)\int_0^{\frac{1}{T}}dt\int d^2x\ \epsilon^{\mu\nu\rho}A_{\mu}\partial_{\nu}A_{\rho}.
\eea 
When we take the zero temperature limit or let $\tanh\big(|m|/(2T)\big)$ approach to one, the non-gauge invariant Abelian Chern-Simons term also consistently goes back to the case of zero temperature.

We can find that the Abelian Chern-Simons generates a temperature dependent term under a large gauge transformation. Hence, the coefficient of the Chern-Simons term cannot be a discrete value. Hence, we cannot use the combination of the effective action and the non-interacting Dirac fermion theory to form a gauge invariant theory at a non-vanishing temperature \cite{Fosco:1997vu}. The issue appears in the effective action due to the perturbative expansion. We will precisely demonstrate the issue from a 0+1 dimensional non-interacting Dirac fermion theory.

\subsection{Parity Anomaly and 0+1 Dimensional Dirac Fermion Theory}
Because the parity anomaly in a 0+1 dimensional non-interacting Dirac fermion theory can be exactly obtained, we introduce the parity anomaly from the theory in this section. Before we show the parity anomaly in the 0+1 dimensional non-interacting Dirac fermion theory, we review necessary mathematical techniques for the infinite product representation of $\cosh(x)$.

\subsubsection{The Infinite Product Representation of $\cosh(x)$}
We start from: 
\bea
\cos(x)&=&\frac{e^{ix}+e^{-ix}}{2}=\frac{1}{2}\lim_{n\rightarrow\infty}\bigg\lbrack\bigg(1+\frac{ix}{n}\bigg)^n+\bigg(1-\frac{ix}{n}\bigg)^n\bigg\rbrack
\nn\\
&=&\frac{1}{2}\lim_{n\rightarrow\infty}\sum_{k=0}^{n}\bigg\lbrack C^n_k\bigg(\frac{ix}{n}\bigg)^k+(-1)^kC^n_k\bigg(\frac{ix}{n}\bigg)^k\bigg\rbrack=\lim_{n\rightarrow\infty}\sum_{k=0}^{\frac{n-1}{2}}(-1)^kC^n_{2k}\frac{x^{2k}}{n^{2k}},
\nn\\
\eea
in which we used 
\bea
e^x=\lim_{n\rightarrow\infty}\bigg(1+\frac{x}{n}\bigg)^n.
\eea
Now we write the above series by an infinite product. We first define
\bea
x\equiv n\tan\theta,
\eea
then we have:
\bea
1+i\frac{x}{n}=1+i\tan\theta=\sec\theta\cdot e^{i\theta}, \qquad 1-i\frac{x}{n}=1-i\tan\theta=\sec\theta\cdot e^{-i\theta}.
\eea
Therefore ,we can deduce: 
\bea
\frac{1}{2}\bigg\lbrack \bigg(1+i\frac{x}{n}\bigg)^n+\bigg(1-i\frac{x}{n}\bigg)^n\bigg\rbrack=\frac{1}{2}\sec^n\theta\cdot(e^{in\theta}+e^{-in\theta})=\sec^n\theta\cdot\cos(n\theta).
\eea
Because we have $\cos(n\theta)=0$ when $n\theta=(k-1/2)\pi$, where $k$ is an integer, we find:
\bea
\cos(x)&\propto&\lim_{n\rightarrow\infty}\prod_{k=1}^{\frac{n-1}{2}}\bigg\lbrack x-n\tan\bigg(\frac{(k-\frac{1}{2})\pi}{n}\bigg)\bigg\rbrack\cdot\bigg\lbrack x+n\tan\bigg(\frac{k-\frac{1}{2}\pi}{n}\bigg)\bigg\rbrack
\nn\\
&=&\lim_{n\rightarrow\infty}\prod_{k=1}^{\frac{n-1}{2}}\bigg\lbrack x^2-n^2\tan^2\bigg(\frac{(k-\frac{1}{2})\pi}{n}\bigg)\bigg\rbrack.
\nn\\
\eea
Thus, we get:
\bea
\cos(x)=\lim_{n\rightarrow\infty}\prod_{k=1}^{\frac{n-1}{2}}\bigg\lbrack1-\frac{x^2}{n^2\tan^2\bigg(\frac{(k-\frac{1}{2})\pi}{n}\bigg)}\bigg\rbrack=\prod_{n=1}^{\infty}\bigg(1-\frac{x^2}{(n-\frac{1}{2})^2\pi^2}\bigg).
\eea
Then we can use the relation $\cosh(x)=\cos(ix)$ to obtain the infinite product representation of $\cosh(x)$
\bea
\cosh x=\prod_{n=1}^{\infty}\bigg(1+\frac{x^2}{(n-\frac{1}{2})^2\pi^2}\bigg).
\eea

\subsubsection{Exact Effective Action and Parity Anomaly}
The action of the one dimensional non-interacting massive Dirac fermion theory in the Euclidean space is
\bea
S_{\mathrm{fermionome}}=\int dt\ \psi^{\dagger}(\partial_t-iA+m)\psi
\eea
with the anti-periodic boundary
\bea
\psi(0)=-\psi(\beta).
\eea
The effective action of the one dimensional non-interacting massive Dirac fermion theory in the Euclidean space is
\bea
S_{\mathrm{effome}}=\ln\bigg(\frac{\det(\partial_t-iA+m)}{\det(\partial_t+m)}\bigg).
\eea
The eigenfunctions are
\bea
\psi_n(t)=\exp\bigg((\Lambda_n-m)t+i\int^t dt^{\prime}\ A(t^{\prime})\bigg),
\eea
where $\Lambda_n$ is a constant for each $n$.
By the anti-periodic boundary condition, we obtain
\bea
\Lambda_n=m-i\frac{a}{\beta}+\frac{(2n-1)\pi i}{\beta},
\eea
where
\bea
a\equiv\int_0^{\beta}dt\ A(t).
\eea
We first compute:
\bea
\frac{\det(\partial_t-iA+m)}{\det(\partial_t+m)}=\prod_{n=-\infty}^{\infty}\bigg(\frac{m-i\frac{a}{\beta}+\frac{(2n-1)\pi i}{\beta}}{m+\frac{(2n-1)\pi i}{\beta}}\bigg)=\frac{\cosh\big(\frac{\beta m}{2}-i\frac{a}{2}\big)}{\cosh\big(\frac{\beta m}{2}\big)},
\eea
in which we used 
\bea
\cosh(x)=\prod_{n=1}^{\infty}\bigg(1+\frac{x^2}{\pi^2(n-\frac{1}{2})^2}\bigg)
\eea
in the last equality. The last equality can be shown:
\bea
\frac{\cosh\big(\frac{\beta m}{2}-i\frac{a}{2}\big)}{\cosh\big(\frac{\beta m}{2}\big)}&=&\prod_{n=1}^{\infty}\frac{1+\frac{(\frac{\beta m}{2}-i\frac{a}{2})^2}{\pi^2(n-\frac{1}{2})^2}}{1+\frac{\frac{\beta^2m^2}{4}}{\pi^2(n-\frac{1}{2})^2}}
=\prod_{n=1}^{\infty}\frac{\frac{4\pi^2(n-\frac{1}{2})^2}{\beta^2}+\frac{(\beta m-ia)^2}{\beta^2}}{\frac{4\pi^2(n-\frac{1}{2})^2}{\beta^2}+m^2}
\nn\\
&=&\prod_{n=1}^{\infty}\frac{\frac{(\beta m-ia)^2}{\beta^2}+\frac{\pi^2(2n-1)^2}{\beta^2}}{m^2+\frac{\pi^2(2n-1)^2}{\beta^2}}
\nn\\
&=&\prod_{n=1}^{\infty}\frac{\bigg(\frac{\beta m-ia}{\beta}+i\frac{\pi(2n-1)}{\beta}\bigg)\bigg(\frac{\beta m-ia}{\beta}-\frac{i\pi(2n-1)}{\beta}\bigg)}{\bigg(m+i\frac{\pi(2n-1)}{\beta}\bigg)\bigg(m-i\frac{\pi(2n-1)}{\beta}\bigg)}
\nn\\
&=&\prod_{n=-\infty}^{n=\infty}\frac{m-i\frac{a}{\beta}+\frac{(2n-1)\pi i}{\beta}}{m+\frac{(2n-1)\pi i}{\beta}}.
\eea
From the above computation, we obtain the exact effective action of the one dimensional non-interacting massive Dirac fermion theory in the Euclidean space 
\bea
S_{\mathrm{effome}}=\ln\bigg\lbrack\cos\bigg(\frac{a}{2}\bigg)-i\tanh\bigg(\frac{\beta m}{2}\bigg)\sin\bigg(\frac{a}{2}\bigg)\bigg\rbrack.
\eea
When we take the zero temperature limit and consider leading order with respect to a number of the gauge field, this exact effective action becomes
\bea
-\frac{i}{2}\frac{m}{|m|}\int dt\ A(t).
\eea
Note that the effective action $S_{\mathrm{effome}}$ is not modified by any temperature dependent term from the large gauge transformation
\bea
a\rightarrow a+2\pi.
\eea
Although the temperature dependent term from the perturbation at the leading order with respect to a number of gauge field is not gauge invariant under the large transformation, the issue of the large gauge transformation can disappear when we do the perturbation to all orders \cite{Deser:1997nv}. Now we also find that the gauge invariant theory can be found by combining the exact effective action $S_{\mathrm{effome}}$ and the 0+1 dimensional non-interacting Dirac fermion theory \cite{Fosco:1997vu}. In the 0+1 dimensional Dirac fermion theory, the issue of the large gauge transformation at a finite temperature should be clear.

\section{A 2+1 Dimensional Duality Web for the Flat Background}
\label{3}
We first introduce a particle-vortex duality of bosons in 2+1 dimensions for the flat background at the IR limit \cite{Peskin:1977kp}, then introduce a conjecture for equivalence between one boson system and one fermion system at the IR limit in 2+1 dimensions for including the particle vortex duality of bosons \cite{Karch:2016sxi, Seiberg:2016gmd} and derive the particle-vortex duality of fermions \cite{Metlitski:2015eka} in a 2+1 dimensional duality web \cite{Karch:2016sxi, Seiberg:2016gmd}. 

\subsection{A 2+1 Dimensional Particle-Vortex Duality for Bosons}
We start from the action of the XY-model
\bea
S_{\mathrm{XY}}=\int d^3x\ \bigg(\frac{1}{2}|(\partial_{\mu}+iA_{\mu})\phi|^2-V(\phi^{\dagger}\phi)\bigg),
\eea
where $A_{\mu}$ is a U(1) background gauge field, $\phi$ is a scalar field and $V(\phi^{\dagger}\phi)$ is a potential of the scalar field. We choose the vacuum expectation value for the complex scalar field $\phi=v\cdot\exp(i\theta)$, where $v$ is a constant, to minimize the potential $V$. Thus, the action becomes
\bea
S_{\mathrm{XY}}\rightarrow\int d^3x\ \frac{1}{2}v^2\bigg(\partial_{\mu}\theta+A_{\mu}\bigg)^2.
\eea
Now we introduce the auxiliary field $\xi_{\mu}$ to write an alternative form
\bea
\int d^3x\ \bigg(-\frac{1}{2v^2}\xi_{\mu}\xi^{\mu}+\xi^{\mu}(\partial_{\mu}\theta+A_{\mu})\bigg).
\eea
Then we use the field redefinition $\theta\equiv\theta_{\mathrm{smooth}}+\theta_{\mathrm{vortex}}$, where $\theta_{\mathrm{smooth}}$ is a smooth configuration of the $\theta$ field and $\theta_{\mathrm{vortex}}$ is not globally defined, to get
\bea
\int d^3x\ \bigg(-\frac{1}{2v^2}\xi_{\mu}\xi^{\mu}+\xi^{\mu}(\partial_{\mu}\theta_{\mathrm{smooth}}+\partial_{\mu}\theta_{\mathrm{vortex}}+A_{\mu})\bigg).
\eea
Now we integrate out the field $\theta_{\mathrm{smooth}}$ to obtain $\partial_{\mu}\xi^{\mu}=0$, which is equivalent to using
\bea
\xi^{\mu}=\frac{1}{2\pi}\epsilon^{\mu\nu\rho}\partial_{\nu}a_{\lambda}.
\eea
Hence, we obtain
\bea
&&\int d^3x\ \bigg(-\frac{1}{4v^2}f_{\mu\nu}f^{\mu\nu}+\frac{1}{2\pi}\epsilon^{\mu\nu\lambda}\partial_{\nu}a_{\lambda}\big(\partial_{\mu}\theta_{vortex}+A_{\mu}
\big)\bigg)
\nn\\
&=&
\int d^3x\ \bigg(-\frac{1}{4v^2}f_{\mu\nu}f^{\mu\nu}+ a_{\mu}j^{\mu}_{vortex}+\frac{1}{2\pi}A_{\mu}\epsilon^{\mu\nu\lambda}\partial_{\nu}a_{\lambda}\bigg),
\eea
where $f_{\mu\nu}\equiv\partial_{\mu}a_{\nu}-\partial_{\nu}a_{\mu}$ and $j_{\mathrm{vortex}}^{\lambda}\equiv\big(1/(2\pi)\big)\epsilon^{\lambda\mu\nu}\partial_{\mu}\partial_{\nu}
\theta_{\mathrm{vortex}}$. Because the field $\theta_{\mathrm{vortex}}$ is not globally defined, this term does not vanish. The vortex current $j_{\mathrm {vortex}}$ coupled to the new gauge field $a_{\mu}$, which also promotes another vortex scalar field $\Phi$ to couple the gauge field $a_{\mu}$. Thus, the action of the particle-vortex dual theory or the Abelian-Higgs theory is given by 
\bea
S_{\mathrm{AH}}=\int d^3x\ \bigg(-\frac{1}{4v^2}f_{\mu\nu}f^{\mu\nu}+\frac{1}{2}|(\partial_{\mu}+ia_{\mu})\Phi|^2-V(\Phi^{\dagger}\Phi)+\frac{1}{2\pi}A_{\mu}\epsilon^{\mu\nu\lambda}\partial_{\nu}a_{\lambda}\bigg).
\eea

\subsection{The 2+1 Dimensional Duality Web}
We first present the conjecture for the equivalence between a boson system and a fermion system at the IR limit \cite{Karch:2016sxi}. Then we use the conjecture to derive the dualities, which are related to the particle-vortex duality of bosons and the particle-vortex duality of fermions at the IR limit \cite{Karch:2016sxi}.

\subsubsection{Conjecture of the 2+1 Dimensional Duality Web}
The conjecture of the 2+1 dimensional duality web is
\bea
Z_{\mathrm{fermion}}[A]\exp\bigg(-\frac{i}{2}S_{\mathrm{CS}}[A]\bigg)=\int D\phi Da\ \exp\big(i S_{\mathrm{scalar}}[\phi; a]+iS_{\mathrm{CS}}[a]+iS_{\mathrm{BF}}[a; A]\big),
\nn\\
\eea
in which the action of the massless Dirac fermion theory is
\bea
S_{\mathrm{fermion}}[\psi; A]\equiv\int d^3x\ i\bar{\psi}\gamma^{\mu}(\partial_{\mu}+iA_{\mu})\psi,
\eea
the partition function of the massless Dirac fermion is
\bea
Z_{\mathrm{fermion}}[A]\equiv\int D\psi\ \exp\big(iS_{\mathrm{fermion}}[\psi ;A]\big),
\eea
the action of the scalar field theory is
\bea
S_{\mathrm{scalar}}[\phi; A]\equiv\int d^3x\ \big(|(\partial_{\mu}+iA_{\mu})\phi|^2-\lambda|\phi|^4\big),
\eea
the action of the BF theory is
\bea
S_{\mathrm{BF}}[a ; A]\equiv\frac{1}{2\pi}\int d^3x\ \epsilon^{\mu\nu\rho}a_{\mu}\partial_{\nu}A_{\rho}.
\eea
The action of the BF theory also has the property
\bea
S_{\mathrm{BF}}[a; A]=S_{\mathrm{BF}}[A; a]
\eea
up to a boundary term. The conjecture is based on cancellation of the parity anomaly and consistency of the $a$-charge and the $A$-charge. 

The massless Dirac fermion is not gauge invariant from the one-loop correction. Hence, the addition of the Abelian Chern-Simons theory with the level one-half is necessary for restoring the gauge symmetry. This only shows that the fermion theory is a gauge invariant theory as in the boson theory.

Now we mention the consistency of charges. A monopole operator of  the dynamical gauge field $a$ has one $a$-charge and one $A$-charge and a monopole operator of the complex scalar field $\phi$ has the vanishing $A$-charge and also has one-half $a$-charge. Thus, we can use the monopole operators of the complex scalar field and the monopole operators of the dynamical gauge field $a$ to construct a gauge invariant operator with one $A$-charge by combining the monopole operator of the complex scalar field and the monopole operator of the dynamical gauge field with vanishing $a$-charge. The gauge invariant operator has the same $A$-charge as the fermion field.

Now we discuss about a valid energy scale of the conjecture. The kinetic term of the Abelian gauge theory should be truncated at the IR limit or $e^2\rightarrow\infty$, where $e$ is a gauge coupling constant.  This conjecture of the 2+1 dimensional duality web also does not have the kinetic term so the conjecture  is better to be formulated at the IR limit.

The charge density of the complex scalar field comes from the equation of motion of $a_0$:
\bea
\rho_{\mathrm{scalar}}+\frac{f_{12}}{2\pi}=0, \qquad f_{12}\equiv\partial_1a_2-\partial_2 a_1,
\eea
where $\rho_{\mathrm{scalar}}$ is charge density of the complex scalar field. We already turned off the background gauge field $A$ in the above equation of motion of $a_0$. This also implies that one unit flux turns on one complex scalar field.

The conjecture of the 2+1 dimensional duality web can be rewritten for convenience of deriving other dualities:
\bea
&&\int D\psi DA\ \exp\bigg(iS_{\mathrm{fermion}}[\psi ; A]-\frac{i}{2}S_{\mathrm{CS}}[A]-iS_{\mathrm{BF}}[A ;C]\bigg)
\nn\\
&=&\int D\phi Df\ \exp\bigg(i S_{\mathrm{scalar}}[\phi; C+df]+iS_{\mathrm{CS}}[C+df]\bigg)
\nn\\
&=&e^{iS_{\mathrm{CS}}[C]}\int Df\ Z_{\mathrm{scalar}}[C+df],
\nn\\
Z_{\mathrm{scalar}}[A]&\equiv&\int D\phi\ \exp\big(iS_{\mathrm{scalar}}[\phi ;A]\big),
\eea
in which we equivalently use $dA=dC$ to obtain $A=C+df$ in the boson theory. The total derivative term is ignored in the second equality. 

The charge density of the fermion field comes from the equation of the motion of the zeroth component of the gauge field $A_0$ by turning off the background gauge field $C$:
\bea
\rho_{\mathrm{fermion}}-\frac{1}{2}\frac{F_{12}}{2\pi}=0, \qquad F_{12}\equiv\partial_1 A_2-\partial_2 A_1,
\eea
where $\rho_{\mathrm{fermion}}$ is the charge density of the fermion field. Thus, this result shows that the fermion charge is one half $Q_{\mathrm{fermion}}=1/2$ when single monopole appears.

We can also use the time reversed operation, which flips the sign of the Chern-Simons and BF terms, to act on the conjecture to get
\bea
Z_{\mathrm{fermion}}[A]e^{\frac{i}{2}S_{\mathrm{CS}}[A]}=\int D\phi Da\ \exp\bigg(iS_{\mathrm{scalar}}[\phi ; a]-iS_{\mathrm{CS}}[a]-iS_{\mathrm{BF}}[a ; A]\bigg)
\eea
and
\bea
&&\int D\psi DA\ \exp\bigg(iS_{\mathrm{fermion}}[\psi ; A]+\frac{i}{2}S_{\mathrm{CS}}[A]+iS_{\mathrm{BF}}[A ;C]\bigg)
\nn\\
&=&e^{-iS_{\mathrm{CS}}[C]}\int Df\ Z_{\mathrm{scalar}}[C+df].
\nn\\
\eea

\subsection{Boson-Boson Duality}
Now we derive a 2+1 dimensional boson-boson duality from the conjecture of the 2+1 dimensional duality web and begin from
\bea
&&e^{-iS_{\mathrm{CS}}[C]}\int D\psi DA\ \exp\bigg(iS_{\mathrm{fermion}}[\psi ; A]-\frac{i}{2}S_{\mathrm{CS}}[A]-iS_{\mathrm{BF}}[A ;C]\bigg)
\nn\\
&=&\int Df\ Z_{\mathrm{scalar}}[C+df].
\nn\\
\eea
We first add the BF term to obtain
\bea
&&\int Da\ \exp\bigg(-iS_{\mathrm{CS}}[a]+iS_{\mathrm{BF}}[a ; A]\bigg)\int D\psi D\tilde{a}
\nn\\
&&\times\exp\bigg(iS_{\mathrm{fermion}}[\psi ; \tilde{a}]-\frac{i}{2}S_{\mathrm{CS}}[\tilde{a}]-iS_{\mathrm{BF}}[\tilde{a} ;a]\bigg)
\nn\\
&=&\int Da\ \exp\bigg(iS_{\mathrm{BF}}[a ; A]\bigg)\int Df\ Z_{\mathrm{scalar}}[a+df],
\eea
then we integrate out the dynamical gauge field $a$ in the fermion theory, which is equivalent to using $a=A-\tilde{a}+dg$, to obtain
\bea
&&\int D\psi D\tilde{a}\ \exp\bigg(iS_{\mathrm{fermion}}[\psi ; \tilde{a}]
+\frac{i}{2}S_{\mathrm{CS}}[\tilde{a}]-iS_{\mathrm{BF}}[\tilde{a} ;A]+iS_{\mathrm{CS}}[A]\bigg)
\nn\\&=&\int Da\ \exp\bigg(iS_{\mathrm{BF}}[a ; A]\bigg)\int Df\ Z_{\mathrm{scalar}}[a+df]
\eea
by computing:
\bea
-iS_{\mathrm{BF}}[\tilde{a}; a]&=&-iS_{\mathrm{BF}}[\tilde{a}; A-\tilde{a}+dg]=-iS_{\mathrm{BF}}[\tilde{a}; A]+2iS_{\mathrm{CS}}[\tilde{a}],
\nn\\
iS_{\mathrm{BF}}[a; A]&=&iS_{\mathrm{BF}}[A-\tilde{a}+dg; A]=2iS_{\mathrm{CS}}[A]-iS_{\mathrm{BF}}[\tilde{a}; A],
\nn\\
-iS_{\mathrm{CS}}[a]&=&-iS_{\mathrm{CS}}[A-\tilde{a}+dg]=-iS_{\mathrm{CS}}[A]-iS_{\mathrm{CS}}[\tilde{a}]+iS_{\mathrm{BF}}[A; \tilde{a}-dg]
\nn\\
&=&-iS_{\mathrm{CS}}[A]-iS_{\mathrm{CS}}[\tilde{a}]+iS_{\mathrm{BF}}[A; \tilde{a}],
\eea
\bea
-\frac{i}{2}S_{\mathrm{CS}}[\tilde{a}]-iS_{\mathrm{BF}}[\tilde{a}, a]-iS_{\mathrm{CS}}[a]+iS_{\mathrm{BF}}[a; A]&=&\frac{i}{2}S_{\mathrm{CS}}[\tilde{a}]-iS_{\mathrm{BF}}[\tilde{a}; A]+iS_{\mathrm{CS}}[A]
\nn\\
\eea
through:
\bea
S_{\mathrm{BF}}[\tilde{a}, \tilde{a}]&=&-S_{\mathrm{BF}}[-\tilde{a}, \tilde{a}]=-S_{\mathrm{BF}}[\tilde{a}, -\tilde{a}]=2S_{\mathrm{CS}}[\tilde{a}],
\nn\\
S_{\mathrm{BF}}[\tilde{a}, A+B]&=&S_{\mathrm{BF}}[\tilde{a}, A]+S_{\mathrm{BF}}[\tilde{a}, B],
\nn\\
S_{\mathrm{CS}}[A]&=&S_{\mathrm{CS}}[-A],
\nn\\
S_{\mathrm{CS}}[A+B]&=&S_{\mathrm{CS}}[A]+S_{\mathrm{CS}}[B]-S_{\mathrm{BF}}[A ;B].
\eea
Then we use the time reversed partition function
\bea
&&\int D\psi DA\ \exp\bigg(iS_{\mathrm{fermion}}[\psi ; A]+\frac{i}{2}S_{\mathrm{CS}}[A]-iS_{\mathrm{BF}}[A ;C]\bigg)
\nn\\
&=&e^{-iS_{\mathrm{CS}}[C]}\int Df\ Z_{\mathrm{scalar}}[-C+df]
\nn\\
\eea
to show
\bea
\int Df\ Z_{\mathrm{scalar}}[-A+df]=\int Da\ \exp\bigg(iS_{\mathrm{BF}}[a ; A]\bigg)\int Df\ Z_{\mathrm{scalar}}[a+df].
\eea
Since the gauge transformation of the dynamical gauge field $a$ is $d\lambda$, where $\lambda$ is a gauge parameter, and shifting of the dynamical gauge field $a$ ($a\rightarrow a-df$) does not modify the BF term, we obtain the equivalence
\bea
\int Df\ Z_{\mathrm{scalar}}[-A+df]\sim\int Da\ \exp\bigg(iS_{\mathrm{BF}}[a ; A]\bigg)Z_{\mathrm{scalar}}[a].
\eea
The path integral of the field $f$ can also be absorbed by the field redefinition of the complex scalar field $\phi\rightarrow\phi\cdot\exp(-if)$. Thus, we finally obtain the boson-boson duality
\bea
Z_{\mathrm{scalar}}[-A]\sim\int Da\ \exp\bigg(iS_{\mathrm{BF}}[a ; A]\bigg)Z_{\mathrm{scalar}}[a].
\eea
We remind that the particle-vortex duality of the bosons shows that the XY-model 
\bea
S_{\mathrm{XY}}=\int d^3x\ \bigg(\frac{1}{2}|(\partial_{\mu}+iA_{\mu})\phi|^2-V(\phi^{\dagger}\phi)\bigg)
\eea
duals to the Abelian-Higgs theory
\bea
S_{\mathrm{AH}}=\int d^3x\ \bigg(-\frac{1}{4v^2}f_{\mu\nu}f^{\mu\nu}+\frac{1}{2}|(\partial_{\mu}+ia_{\mu})\Phi|^2-V(\Phi^{\dagger}\Phi)+\frac{1}{2\pi}A_{\mu}\epsilon^{\mu\nu\lambda}\partial_{\nu}a_{\lambda}\bigg)
\eea
at the IR limit.
When we do the identification:
\bea
V(\phi^{\dagger}\phi)=\lambda |\phi|^4, \qquad V(\Phi^{\dagger}\Phi)=\lambda|\Phi|^4
\eea
and the kinetic term of the Abelian gauge field 
\bea
-\frac{1}{4v^2}f_{\mu\nu}f^{\mu\nu}
\eea
should be truncated under the IR limit, we find that the boson-boson duality is equivalent to the particle-vortex duality of bosons. Hence, this means that the conjecture of the 2+1 dimensional duality web leads a provable duality at the IR limit. This should give quite reliable evidences to the conjecture of the 2+1 dimensional duality web. However, beginning from the particle-vortex duality of bosons cannot show the conjecture of the 2+1 dimensional duality web. Hence, the conjecture of the 2+1 dimensional duality web was not proved yet.

\subsection{Fermion-Fermion Duality}
We start from the conjecture of the 2+1 dimensional duality web
\bea
&&Z_{\mathrm{fermion}}[C]
\nn\\
&=&\int D\phi D\tilde{a}\ \exp\big(i S_{\mathrm{scalar}}[\phi; \tilde{a}]+iS_{\mathrm{CS}}[\tilde{a}]+iS_{\mathrm{BF}}[\tilde{a}; C]\big)\exp\bigg(\frac{i}{2}S_{\mathrm{CS}}[C]\bigg),
\nn\\
\eea
where $C\equiv 2a+A$. The set of the background $C$ by letting the theory being gauge invariant with a spin structure and results of integration of the dynamical gauge fields should obey the Dirac quantization condition, which is $\int dC=2\pi$. Then we add the BF coupling to the conjecture of the 2+1 dimensional duality web to obtain
\bea
&&\int Da\ Z_{\mathrm{fermion}}[C]\cdot\exp\bigg(\frac{i}{2}S_{\mathrm{BF}}[C; A]\bigg)
\nn\\
&=&\int D\phi D\tilde{a}Da\ \exp\big(i S_{\mathrm{scalar}}[\phi; \tilde{a}]+iS_{\mathrm{CS}}[\tilde{a}]+iS_{\mathrm{BF}}[\tilde{a}; C]\big)\exp\bigg(\frac{i}{2}S_{\mathrm{CS}}[C]\bigg)
\nn\\
&&\times\exp\bigg(\frac{i}{2}S_{\mathrm{BF}}[C; A]\bigg).
\nn\\
\eea
We integrate out the dynamical gauge field $a$ in the boson theory and it is equivalent to using $dC=-(dA+2d\tilde{a})$ or $C=-A-2\tilde{a}+dg$. Thus, we get
\bea
&&\int D\phi D\tilde{a}Da\ \exp\big(i S_{\mathrm{scalar}}[\phi; \tilde{a}]+iS_{\mathrm{CS}}[\tilde{a}]+iS_{\mathrm{BF}}[\tilde{a}; C]\big)\exp\bigg(\frac{i}{2}S_{\mathrm{CS}}[C]\bigg)
\nn\\
&&\times\exp\bigg(\frac{i}{2}S_{\mathrm{BF}}[C; A]\bigg)
\nn\\
&=&\int D\phi D\tilde{a}\ \exp\big(i S_{\mathrm{scalar}}[\phi; \tilde{a}]-iS_{\mathrm{CS}}[\tilde{a}]-iS_{\mathrm{BF}}[\tilde{a}; A]-\frac{i}{2}S_{\mathrm{CS}}[A]\big)
\nn\\
\eea
by computing:
\bea
iS_{\mathrm{BF}}[\tilde{a}; C]&=&iS_{\mathrm{BF}}[\tilde{a}, -A-2\tilde{a}+dg]=-iS_{\mathrm{BF}}[\tilde{a}, A]-4S_{\mathrm{CS}}[\tilde{a}],
\nn\\
\frac{i}{2}S_{\mathrm{CS}}[C]&=&\frac{i}{2}S_{\mathrm{CS}}[-A-2\tilde{a}+dg]=\frac{i}{2}S_{\mathrm{CS}}[A]+2iS_{\mathrm{CS}}[\tilde{a}]+iS_{\mathrm{BF}}[\tilde{a}; A],
\nn\\
\frac{i}{2}S_{\mathrm{BF}}[C; A]&=&\frac{i}{2}S_{\mathrm{BF}}[-A-2\tilde{a}+dg, A]=-iS_{\mathrm{CS}}[A]-iS_{\mathrm{BF}}[\tilde{a}; A],
\eea
\bea
iS_{\mathrm{CS}}[\tilde{a}]+iS_{\mathrm{BF}}[\tilde{a}, C]+\frac{i}{2}S_{\mathrm{CS}}[C]+\frac{i}{2}S_{\mathrm{BF}}[C; A]=-iS_{\mathrm{CS}}[\tilde{a}]-iS_{\mathrm{BF}}[\tilde{a}; A]-\frac{i}{2}S_{\mathrm{CS}}[A].
\nn\\
\eea
Now we use
\bea
Z_{\mathrm{fermion}}[A]\cdot\exp\bigg(\frac{i}{2}S_{\mathrm{CS}}[A]\bigg)=\int D\phi Da\ \exp\bigg(iS_{\mathrm{scalar}}[\phi; a]-iS_{\mathrm{CS}}[a]-iS_{\mathrm{BF}}[a;A]\bigg)
\nn\\
\eea
to obtain:
\bea
Z_{\mathrm{fermion}}[A]&=&\int D\psi Da\ \exp\bigg(i S_{\mathrm{fermion}}[\psi; 2a+A]+\frac{i}{2}S_{\mathrm{BF}}[2a+A; A]\bigg)
\nn\\
&=&
\int D\psi Da\ \exp\bigg(i S_{\mathrm{fermion}}[\psi; 2a+A]+iS_{\mathrm{BF}}[a; A]+iS_{\mathrm{CS}}[A]\bigg).
\nn\\
\eea
Hence, we show one example to demonstrate that the conjecture of the 2+1 dimensional duality web can lead other dualities. We also mention that the fermion-fermion duality is also the particle-vortex duality of fermions at global level. Note that the similar way can also lead other dualities to form the 2+1 dimensional duality web in the flat background \cite{Karch:2016sxi, Seiberg:2016gmd, Ma:2016yas, Ma:2017mpb}. 

\section{The 2+1 Dimensional Duality Web in the Spin$_c$ Manifold}
\label{4}
We first review the spin manifold and the spin$_c$ manifold and clearly find differences between these two manifolds. Then we show why the 2+1 dimensional duality web \cite{Karch:2016sxi, Seiberg:2016gmd} is problematic from the spin structure on the spin manifold \cite{Karch:2016sxi, Seiberg:2016gmd} and the spin$_c$ manifold should be useful in the study of the 2+1 dimensional duality web \cite{Seiberg:2016gmd}.

\subsection{Review of the Spin Manifold}
An $n$-dimensional oriented Riemannian manifold $M$ with the spin structure $P$ is the spin manifold. The spin structure is defined by the the canonical diagram:
\[
\begin{tikzcd}
P\times \mathrm{Spin}(n) \arrow{r}{} \arrow[swap]{d}{} & P \arrow{d}{}\arrow[swap]{dr}{}  \\ 
P_{\mathrm{SO}}M\times \mathrm{SO}(n) \arrow{r}{}& P_{\mathrm{SO}}M \arrow{r}{} & M\qquad 
\end{tikzcd}
\]
This is the $\mathrm{Spin}(n)$-principle bundle which doubly covers the bundle of the oriented tangent frames $P_{\mathrm{SO}}M$ of the manifold $M$ such that the canonical diagram commutes. 

Now we define the principle bundle. The principle bundle formalizes the Cartesian product $X\times G$, where $X$ is a space and $G$ is a group. The equivalent way as the Cartesian product, the principle bundle $B$ is equipped with:

1. An action of the group $G$ on the principle bundle $B$, analogous to the the product space $(X, g)h=(X, gh)$.

2. A projection onto the space $X$. For the product space, it is just the projection onto the first factor as $(x, G)\mapsto X$.

The spin group $\mathrm{Spin}(n, R)$ is defined by the special orthogonal group $\mathrm{SO}(n, R)$. The special orthogonal group $\mathrm{SO}(n, R)$ is not simply connected. Therefore, the fundamental group of the special orthogonal group shows
\bea
\pi_1\big(\mathrm{SO}(n, R)\big)=Z_2.
\eea
The simply connected double cover is the spin group $\mathrm{Spin}(n, R)$. We list some examples: $\mathrm{Spin}(1)=Z_2$, $\mathrm{Spin}(2)$ is a circle and a double covering of $\mathrm{O}(2)$, $\mathrm{Spin}(3)=\mathrm{SU}(2)=\mathrm{Sp}(1)$, $\mathrm{Spin}(4)=\mathrm{SU}(2)\times \mathrm{SU}(2)$, $\mathrm{Spin}(5)=\mathrm{Sp}(2)$ and $\mathrm{Spin}(6)=\mathrm{SU}(4)$. 

We show one simple example to demonstrate why the spin manifold depends on a choice of the spin structures. The spectrum of the Dirac operator $i(d/dt)$ depends on a choice of the spin structures on the circle $S^1=R/2\pi Z$. Because the frame bundle $P_{\mathrm{SO}}S_1$ is trivial, we can write the trivial spin structure $P=S^1\times \mathrm{Spin}(1)$. The associated spinor bundle is also trivial and it is one dimension. Thus, the spinor is a $C$-valued function on the circle $S^1$. The Fourier analysis shows that the spectrum consists of the eigenvalues $\lambda_k=k$ with the corresponding eigenfunctions $t\mapsto \exp(-ikt)$, where $k\in Z$.

We also choose the other spin structure $\tilde{P}=\big(\lbrack 0, 2\pi\rbrack\times \mathrm{Spin}(1)\big)/S^1$, in which the circle $S^1$ identifies $0$ with $2\pi$ when it exchanges these two elements of $\mathrm{Spin}(1)$. Spinors with respect to the spin structure $\tilde{P}$ no longer correspond to functions on the circle $S^1$, i.e. $2\pi$-periodic functions, but this corresponds to $2\pi$ anti-periodic complex-valued functions as $\psi(t+2\pi)=-\psi(2\pi)$. The eigenvalues are $\lambda_k=k+1/2$, where $k\in Z$, and its corresponding eigenfunctions $t\mapsto\exp\big(-i(k+1/2)\big)t\big)$. Thus, the simple example shows that the eigenvalues of the Dirac operator depends on a choice of the spin structure. 

Now we show that the Chern-Simons theory on an oriented three manifold or the spin manifold also depends on a choice of the spin structures even if the Chern-Simons theory is gauge invariant. The Chern-Simons theory always exists the oriented four manifold $X$ with the boundary manifold $W$. If the oriented three manifold $W$ has a chosen spin structure, the oriented four manifold $X$ can be chosen so that the spin structure of the oriented three manifold $W$ extends over the oriented four manifold $X$. Picking the oriented four manifold $X$, then we can get
\bea
\frac{1}{4\pi}\int_Wd^3x\ \epsilon^{\mu\nu\rho}A_{\mu}\partial_{\nu}A_{\rho}=2\pi\cdot\frac{1}{8}\int_X\ d^4x\ \epsilon^{\mu\nu\rho\sigma}\frac{F_{\mu\nu}}{2\pi}\frac{F_{\rho\sigma}}{2\pi}.
\eea
We can also choose the other oriented four manifold $\tilde{X}$ and the other oriented three manifold $\tilde{W}$ to get
\bea
\frac{1}{4\pi}\int_{\tilde{W}}d^3x\ \epsilon^{\mu\nu\rho}A_{\mu}\partial_{\nu}A_{\rho}=2\pi\cdot\frac{1}{8}\int_{\tilde{X}}\ d^4x\ \epsilon^{\mu\nu\rho\sigma}\frac{F_{\mu\nu}}{2\pi}\frac{F_{\rho\sigma}}{2\pi}.
\eea
Then we glue these manifolds and obtain the oriented four manifold without boundary $X^{\prime}$ to show
\bea
&&\frac{1}{4\pi}\int_Wd^3x\ \epsilon^{\mu\nu\rho}A_{\mu}\partial_{\nu}A_{\rho}-\frac{1}{4\pi}\int_{\tilde{W}}d^3x\ \epsilon^{\mu\nu\rho}A_{\mu}\partial_{\nu}A_{\rho}
\nn\\
&=&2\pi\cdot\frac{1}{8}\int_{X^{\prime}}\ d^4x\ \epsilon^{\mu\nu\rho\sigma}\frac{F_{\mu\nu}}{2\pi}\frac{F_{\rho\sigma}}{2\pi}.
\eea
Because the oriented four manifold $X^{\prime}$ does not have boundary, the difference should be $2\pi Z$. Hence, the Chern-Simons theory on a spin manifold depends on a choice of the spin structure. Because the difference is $2\pi Z$, the difference does not cause non-gauge invariance.

\subsection{Review of the Spin$_c$ Manifold}
The definition of the spin$_c$ manifold is analogous to the definition of the spin manifold. When an oriented Riemannian manifold admits a spin$_c$ structure, the manifold is called the spin$_c$ manifold.

 We first define the group $\mathrm{Spin} _c(n)\equiv\big(\mathrm{Spin}(n)\times \mathrm{U}(1)\big)/\{\pm(1, 1)\}\equiv \mathrm{Spin}(n)\times_{Z_2}\mathrm{U}(1)$. A spin$_c$ structure on the oriented tangent frame $P_{\mathrm{SO}(n)}$ consists of the principle $\mathrm{U}_1$-bundle $P_{\mathrm{U}(1)}$ and also the $\mathrm{Spin}_c(n)$-principle bundle $P_{\mathrm{Spin}_c(n)}$ with the $\mathrm{Spin}_c(n)$ equivalent bundle map $P_{\mathrm{Spin}_c(n)}\rightarrow P_{\mathrm{SO}(n)}\times P_{\mathrm{U}(1)}$. 
 
All oriented smooth manifolds less than five dimensions admit a spin$_c$ structure. The spin manifold also admits the spin$_c$ structure. Hence, the spin$_c$ manifold should be more general than the spin manifold. 

We use the Atiyah-Patodi-Singer theorem without the Dirac fermion field theory \cite{Atiyah:1975jf}:
\bea
\frac{1}{768\pi^2}\int_X d^4x\ \epsilon^{\mu\nu\rho\sigma}R_{\mu\nu}R_{\rho\sigma}+\frac{1}{32\pi^2}\int_X d^4x\ \epsilon^{\mu\nu\rho\sigma}F_{\mu\nu}F_{\rho\sigma}=J,
\eea
in which the manifold $X$ is a four dimensional spin$_c$ manifold, and $J$ is an integer. We also know the relation
\bea
2\pi\bigg(\frac{1}{768\pi^2}\int_X d^4x\ \epsilon^{\mu\nu\rho\sigma}R_{\mu\nu}R_{\rho\sigma}+\frac{1}{32\pi^2}\int_X d^4x\ \epsilon^{\mu\nu\rho\sigma}F_{\mu\nu}F_{\rho\sigma}\bigg)=2S_{\mathrm{grav}}+S_{\mathrm{CS}}.
\eea
Hence, we can obtain
\bea
\pi J=S_{\mathrm{grav}}+\frac{S_{\mathrm{CS}}}{2}.
\eea
 Thus, we can deduce the relation
\bea
2\mbox{CS}_{\mathrm{grav}}=-\mbox{CS}(A), \qquad \mbox{mod}\ 2\pi Z.
\eea
Because we consider the spin$_c$ manifold, the gravitational Chern-Simons term cannot be ignored. Thus, the dependence of the spin structure in the Abelian Chern-Simons theory with the level one does not exist now. 

Now we can obtain the partition function of the Abelian Chern-Simons theory with the level one
\bea
\exp(-2i\Omega_0),
\eea
where $\Omega_0\equiv \mbox{CS}_{\mathrm{grav}}$. We also find the well-defined expression
\bea
\frac{1}{4\pi}\int_M d^3x\ \epsilon^{\mu\nu\rho}A_{\mu}\partial_{\nu}A_{\rho}+\Omega \qquad \mbox{mod}\ 2\pi,
\eea
where $2\Omega_0\equiv\Omega$ and $M$ is a three dimensional oriented manifold with the spin$_c$ structure. Hence, an action of the spin$_c$ version of $\mathrm{U}(1)_1$ or $\mathrm{U}(1)$ with the level l gauge theory is given by
\bea
&&\frac{1}{4\pi}\int_Md^3x\ \epsilon^{\mu\nu\rho}\big(b_{\mu}\partial_{\nu}b_{\rho}+2b_{\mu}\partial_{\nu}A_{\rho}\big)
\nn\\
&=&\frac{1}{4\pi}\int_Md^3x\ \epsilon^{\mu\nu\rho}\bigg(\big(b_{\mu}+A_{\mu}\big)\partial_{\nu}\big(b_{\rho}+A_{\rho}\big)
-A_{\mu}\partial_{\nu}A_{\rho}\bigg),
\nn\\
\eea
where $b$ is a dynamical Abelian gauge field, $A$ is a background spin$_c$ gauge field, and $b+A$ is also a spin$_c$ gauge field. After we integrate out the dynamical gague field $b$, then we obtain the partition function
\bea
\exp(-2i\Omega_0)\exp\bigg(-\frac{i}{4\pi}\int_Md^3x\ \epsilon^{\mu\nu\rho}A_{\mu}\partial_{\nu}A_{\rho}\bigg).
\eea
Therefore, the theory is gauge invariant and does not depend on a choice of the spin structures on the spin$_c$ manifold. 

Finally, we remind that although all oriented manifolds less than five dimensions admit a spin$_c$ gaue field, it does not imply that the spin$_c$ gauge field is the same as the ordinary $\mathrm{U}(1)$ gauge field exactly. The spin$_c$ gauge field is only locally same as the ordinary $\mathrm{U}(1)$ gauge field, but they can be different from a global sector.

\subsection{The Problem of the 2+1 Dimensional Duality Web in the Spin Manifold}
Now we explicitly show why building the 2+1 dimensional duality web in the spin$_c$ manifold is necessary. We begin from the simplified notation of the conjecture of the 2+1 dimensional duality web with the curved background
\bea
i\bar{\psi}\slashed{D}_A\psi\longleftrightarrow |D_a\phi|^2-\lambda|\phi|^4+\frac{1}{4\pi}ada+\frac{1}{2\pi}adA,
\eea
where 
\bea
\slashed{D}_A\equiv \gamma^{\mu}(\partial_{\mu}+iA_{\mu}), \qquad D_A\equiv \partial_{\mu}+iA_{\mu}.
\eea
We also ignore all background terms  (gravitation Chern-Simons term and Abelian Chern-Simons term with the background gauge field) in the simplified notation of the conjecture of the 2+1 dimensional duality web. If we consider the spin$_c$ manifold in the conjecture of the 2+1 dimensional duality web, the background gauge field should be the spin$_c$ background gauge field. 

Now we can add 
\bea
\frac{1}{2\pi}adB-\frac{1}{4\pi}BdB
\eea
to the simplified notation of the conjecture of the 2+1 dimensional duality web and promote the background field $A$ to be the dynamical gauge field $b$
\bea
&&i\bar{\psi}\slashed{D}_b\psi+\frac{1}{2\pi}bdB-\frac{1}{4\pi}BdB
\nn\\
&\longleftrightarrow& |D_a\phi|^2-\lambda|\phi|^4+\frac{1}{4\pi}ada+\frac{1}{2\pi}adb+\frac{1}{2\pi}bdB-\frac{1}{4\pi}BdB,
\eea
then we integrate out the dynamical gauge field $b$ in the right-hand side of the duality \cite{Seiberg:2016gmd}
\bea
&&i\bar{\psi}\slashed{D}_b\psi+\frac{1}{2\pi}bdB-\frac{1}{4\pi}BdB
\nn\\
&\longleftrightarrow& |D_{-B}\phi|^2-\lambda|\phi|^4.
\eea
The integration of the dynamical gauge field $b$ does not alter the ignored background terms in the boson theory of the simplified notation of the 2+1 dimensional duality web, but the integration cannot be directly used in the fermion theory of the simplified notation of the 2+1 dimensional duality web because the ignored terms should affect the result. Now we find that the boson theory of the right-hand side of the duality can be formulated on a non-spin manifold, but the fermion theory or the left-hand side of the duality depends on a choice of the spin structures on the spin manifold \cite{Seiberg:2016gmd}. This means that the duality is inconsistent on a spin manifold \cite{Seiberg:2016gmd}. As we introduced, this problem can be solved when we consider the spin$_c$ manifold or the spin$_c$ gauge field $b$ \cite{Seiberg:2016gmd}.

\section{Global Description of the Quantum Hall Effect}
\label{5}
We review the composite fermion approach of the 2+1 dimensional quantum Hall effect  \cite{Laughlin:1983fy, Zhang:1988wy}. The 2+1 dimensional duality web defined on the spin$_c$ manifold shows the global effect \cite{Seiberg:2016gmd}. We also demonstrate that the 2+1 dimensional quantum Hall system also suffers from the similar issue \cite{Seiberg:2016gmd} and the issue should affect physical results, not a mathematical issue.

\subsection{Mean-Field Theory}
The two dimensional spinless electrons are described by the mean-field theory or the Hamiltonian:
\bea
H_{\mathrm{2e}}&\equiv& K+V\equiv\frac{1}{2m_e}\int d^2r\ \psi_e^{\dagger}(\vec{r})\bigg\lbrack\bigg(-i\nabla-\vec{A}(\vec{r})\bigg)^2+eA_0(\vec{r})\bigg\rbrack\psi_e(\vec{r})
\nn\\
&&+\frac{1}{2}\int d^2r d^2r^{\prime}\frac{e^2}{4\pi\epsilon |\vec{r}-\vec{r}^{\prime}|}\rho(\vec{r})\rho(\vec{r}^{\prime}),
\eea
in which the kinetic energy is defined by
\bea
K\equiv \frac{1}{2m_e}\int d^2r\ \psi_e^{\dagger}(\vec{r})\bigg\lbrack\bigg(-i\nabla-\vec{A}(\vec{r})\bigg)^2+eA_0(\vec{r})\bigg\rbrack\psi_e(\vec{r}),
\eea
the potential energy is defined by
\bea
V\equiv \frac{1}{2}\int d^2r d^2r^{\prime}\frac{e^2}{4\pi\epsilon |\vec{r}-\vec{r}^{\prime}|}\rho(\vec{r})\rho(\vec{r}^{\prime}),
\eea
the electron mass is $m_e$, the electron charge is $e$, and the dielectric constant is $\epsilon$. 

\subsection{Quasi-Particle Creation Operator}
Now we introduce the quasi-particle creation operator $\psi^{\dagger}(\vec{r})$ \cite{Zhang:1988wy}:
\bea
\psi^{\dagger}(\vec{r})&\equiv& \psi_e^{\dagger}(\vec{r})\exp\bigg(i\frac{\theta}{\pi}\int d^2r^{\prime}\ \arg\big(\vec{r}-\vec{r}^{\prime}\big)\rho(\vec{r}^{\prime})\bigg), 
\nn\\
\rho(\vec{r})&\equiv&\psi_e^{\dagger}(\vec{r})\psi_e(\vec{r})=\psi^{\dagger}(\vec{r})\psi(\vec{r}),
\nn\\
\int d^2r\ \rho(\vec{r})&=&1,
\eea
where $\theta$ is a constant and $\arg\big(\vec{r}-\vec{r}^{\prime}\big)$ gives an angle of the vector $\vec{r}-\vec{r}^{\prime}$ and 
\bea
-\pi<\arg\big(\vec{r}-\vec{r}^{\prime}\big)\le\pi.
\eea
 Then we use the quasi-particle creation operator $\psi^{\dagger}(\vec{r})$ to rewrite the kinetic energy $K$ \cite{Zhang:1988wy}:
\bea
-i\nabla\psi_e(\vec{r})&=&-i\nabla\bigg\lbrack\exp\bigg(i\frac{\theta}{\pi}\int d^2r^{\prime}\ \arg\big(\vec{r}-\vec{r}^{\prime}\big)\rho(\vec{r}^{\prime})\bigg)\psi(\vec{r})\bigg\rbrack
\nn\\
&=&\frac{\theta}{\pi}\exp\bigg(i\frac{\theta}{\pi}\int d^2r^{\prime}\ \arg\big(\vec{r}-\vec{r}^{\prime}\big)\rho(\vec{r}^{\prime})\bigg)
\nn\\
&&\times\int d^2 r^{\prime\prime}\ \nabla\big(\arg(\vec{r}-\vec{r}^{\prime\prime})\big)\rho(\vec{r}^{\prime\prime})\psi(\vec{r})
\nn\\
&&-i\exp\bigg(i\frac{\theta}{\pi}\int d^2r^{\prime}\ \arg\big(\vec{r}-\vec{r}^{\prime}\big)\rho(\vec{r}^{\prime})\bigg)\nabla\psi(\vec{r}),
\eea
the computation of $\nabla\big(\arg(\vec{r}-\vec{r}^{\prime\prime})\big)$ is given by:
\bea
\arg(\vec{r}-\vec{r}^{\prime})&\equiv&\theta^{\prime},  \qquad
\tan(\theta^{\prime})\equiv\frac{y}{x}, \qquad
\sec^2(\theta^{\prime}) d\theta^{\prime}=\frac{dy}{x}-\frac{y}{x^2}dx=\frac{x^2+y^2}{x^2}d\theta^{\prime},
\nn\\
\frac{\partial\theta^{\prime}}{\partial x}&=&-\frac{y}{x^2+y^2}, \qquad \frac{\partial\theta^{\prime}}{\partial y}=\frac{x}{x^2+y^2},
\nn\\
\nabla_i\big(\arg(\vec{r}-\vec{r}^{\prime})\big)&=&-\epsilon_{ij}\frac{r_j-r^{\prime}_j}{|\vec{r}-\vec{r}^{\prime}|^2}, \qquad \epsilon_{xy}=-\epsilon_{yx}=1, \qquad \epsilon_{xx}=\epsilon_{yy}=0,
\eea
then we obtain the quantization algebra for the quasi-particle:
\bea
-i\nabla\psi_e(\vec{r})&=&\exp\bigg(i\frac{\theta}{\pi}\int d^2r^{\prime}\ \arg\big(\vec{r}-\vec{r}^{\prime}\big)\rho(\vec{r}^{\prime})\bigg)\big(-i\nabla-e\vec{a}(\vec{r})\big)\psi(\vec{r}),
\nn\\
a_i(\vec{r})&\equiv&\frac{\theta}{\pi e}\epsilon_{ij}\frac{r_j-r^{\prime}_j}{|\vec{r}-\vec{r}^{\prime}|^2}\psi^{\dagger}(\vec{r}^{\prime})\psi(\vec{r}^{\prime}),
\nn\\
U^{-1}\bigg(-i\nabla-\vec{A}(\vec{r})\bigg)U&=&-i\nabla-e\vec{a}(\vec{r}), \qquad U\equiv\exp\bigg(i\frac{\theta}{\pi}\int d^2r^{\prime}\ \arg\big(\vec{r}-\vec{r}^{\prime}\big)\rho(\vec{r}^{\prime})\bigg),
\nn\\
\eea
where $a_i$ is the $i$-th component of the gauge potential $\vec{a}$,
and we also obtain the kinetic energy \cite{Zhang:1988wy}
\bea
K=\int d^2r\ \psi^{\dagger}(\vec{r})\bigg(\frac{1}{2m_e}\big(-i\nabla-\vec{A}(\vec{r})-e\vec{a}(\vec{r})\big)^2+eA_0(\vec{r})\bigg)\psi(\vec{r}).
\eea

Finally, we discuss the quantization of the quasi-particle creation operator $\psi^{\dagger}(\vec{r})$ \cite{Zhang:1988wy}:
\bea
\{\psi_e(\vec{r}), \psi^{\dagger}_e(\vec{r}^{\prime})\}&=&\psi_e(\vec{r})\psi_e^{\dagger}(\vec{r}^{\prime})+\psi_e^{\dagger}(\vec{r}^{\prime})\psi_e(\vec{r})=\delta^2(\vec{r}-\vec{r}^{\prime}),
\eea
which is the quantization algebra for the electronic field,
\bea
\psi_e(\vec{r})\psi_e^{\dagger}(\vec{r}^{\prime})&=&\psi(\vec{r})\psi^{\dagger}(\vec{r}^{\prime})\exp\bigg(i\frac{\theta}{\pi}
\int d^2r^{\prime\prime}\ \big(\arg(\vec{r}-\vec{r}^{\prime\prime})\rho(\vec{r}^{\prime\prime})-\arg(\vec{r}^{\prime}
-\vec{r}^{\prime\prime})\rho(\vec{r}^{\prime\prime})\big)\bigg)
\nn\\
&=&\psi(\vec{r})\psi^{\dagger}(\vec{r}^{\prime})\exp\bigg(i\frac{\theta}{\pi}
\int d^2r^{\prime\prime}\ \arg(\vec{r}-\vec{r}^{\prime})\rho(\vec{r}^{\prime\prime})\bigg)
\nn\\
&=&\psi(\vec{r})\psi^{\dagger}(\vec{r}^{\prime})\exp\bigg(i\frac{\theta}{\pi}\arg(\vec{r}-\vec{r}^{\prime})\int d^2r^{\prime\prime}\rho(\vec{r}^{\prime\prime})\bigg)
\nn\\
&=&\psi(\vec{r})\psi^{\dagger}(\vec{r}^{\prime})\exp\bigg(i\frac{\theta}{\pi}\arg(\vec{r}-\vec{r}^{\prime})\bigg),
\eea
in which the second equality uses
\bea
\arg(\vec{r}-\vec{r}^{\prime\prime})-\arg(\vec{r}^{\prime}-\vec{r}^{\prime\prime})=\arg(\vec{r}-\vec{r}^{\prime}),
\eea
the fourth equality uses
\bea
\int d^2r^{\prime\prime}\ \rho(\vec{r}^{\prime\prime})=1,
\eea
then we also obtain:
\bea
\psi_e^{\dagger}(\vec{r})\psi_e(\vec{r}^{\prime})&=&\psi^{\dagger}(\vec{r}^{\prime})\psi(\vec{r})\exp\bigg(i\frac{\theta}{\pi}\arg(\vec{r}^{\prime}-\vec{r})\bigg)
\nn\\
&=&\psi^{\dagger}(\vec{r}^{\prime})\psi(\vec{r})\exp\bigg(i\frac{\theta}{\pi}\arg(\vec{r}-\vec{r}^{\prime})-i\theta\bigg),
\eea
in which the second equality uses
\bea
\arg(\vec{r}^{\prime}-\vec{r})=\arg(\vec{r}-\vec{r}^{\prime})-\pi,
\eea
and the quantization algebra of the quasi-particle operator $\psi^{\dagger}(\vec{r})$ is given by:
\bea
\psi(\vec{r})\psi^{\dagger}(\vec{r}^{\prime})+\psi^{\dagger}(\vec{r}^{\prime})\psi(\vec{r})e^{-i\theta}=e^{-i\frac{\theta}{\pi}\arg(\vec{r}-\vec{r}^{\prime})}\delta^2(\vec{r}-\vec{r}^{\prime}).
\eea
Thus, we obtain \cite{Zhang:1988wy}:
\bea
\lbrack\psi(\vec{r}), \psi^{\dagger}(\vec{r}^{\prime})\rbrack&=&\delta^2(\vec{r}-\vec{r}^{\prime}), \qquad \theta=(2k+1)\pi,
\nn\\
\{\psi(\vec{r}), \psi^{\dagger}(\vec{r}^{\prime})\}&=&\delta^2(\vec{r}-\vec{r}^{\prime}), \qquad \theta=2k\pi,
\eea
where $k$ is an integer. Thus, we can find that the quasi-particle is a boson field when we choose the parameter $\theta=(2k+1)\pi$ and it is a fermion field when we choose the parameter $\theta=2k\pi$ \cite{Zhang:1988wy}. This also means that the statistics of the quasi-particle depends on the flux or the gauge field $\vec{a}(\vec{r})$ \cite{Zhang:1988wy}.

\subsection{Lagrangian}
Now we want to write a Lagrangian for the mean-field theory \cite{Zhang:1988wy}. We first define a new gauge field $b$ from
\cite{Zhang:1988wy}:
\bea
b(\vec{r})&\equiv& -\epsilon_{ij}\partial_ia_j(\vec{r})
=-\epsilon_{ij}\partial_i\bigg(\frac{\theta}{\pi e}\epsilon_{jk}\int d^2r^{\prime}\ \frac{r_k-r^{\prime}_k}{|\vec{r}-\vec{r}^{\prime}|^2}\psi^{\dagger}(\vec{r}^{\prime})\psi(\vec{r}^{\prime})\bigg)
\nn\\
&=&\frac{\theta}{\pi e}\delta_{ik}\partial_i\bigg(\int d^2r^{\prime}\ \frac{r_k-r_k^{\prime}}{|\vec{r}-\vec{r}^{\prime}|^2}\psi^{\dagger}(\vec{r}^{\prime})\psi(\vec{r}^{\prime})\bigg)
\nn\\
&=&\frac{\theta}{\pi e}\partial_i\bigg(\int d^2r^{\prime}\ \frac{r_i-r_i^{\prime}}{|\vec{r}-\vec{r}^{\prime}|^2}\psi^{\dagger}(\vec{r}^{\prime})\psi(\vec{r}^{\prime})\bigg)
\nn\\
&=&\frac{\theta}{\pi e}\partial_i\bigg(\int d^2r^{\prime}\ \partial_i\big(\ln|\vec{r}-\vec{r}^{\prime}|\big)\psi^{\dagger}(\vec{r}^{\prime})\psi(\vec{r}^{\prime})\bigg)
\nn\\
&=&\frac{2\theta}{e}|\psi(\vec{r})|^2,
\eea
in which we used
\bea
\partial_i\partial_i\ln|\vec{r}-\vec{r}^{\prime}|=2\pi \delta^2(\vec{r}-\vec{r}^{\prime})
\eea
in the last equality. The new gauge field $b$ can deduce the following relation \cite{Zhang:1988wy}:
\bea
\dot{b}(\vec{r})\equiv\partial_tb(\vec{r})=-\epsilon_{ij}\partial_i\dot{a}_j=\frac{2\theta}{e}\frac{\partial}{\partial t}|\psi(\vec{r})|^2.
\eea
If we define the current field $j^{\mu}$ as:
\bea
j^0\equiv |\psi(\vec{r})|^2, \qquad \partial_{\mu}j^{\mu}=0,
\eea 
the relations of the new gauge field $b$ can be obtained from the Lagrangian \cite{Zhang:1988wy}
\bea
L_{\mathrm{aj}}=\frac{e^2}{4\theta}\epsilon^{\mu\nu\rho}a_{\mu}\partial_{\nu}a_{\rho}+ea_{\mu}j^{\mu}.
\eea
The variation of the gauge field $a_{\mu}$ gives 
\bea
\frac{e^2}{2\theta}\epsilon^{\mu\nu\rho}\partial_{\nu}a_{\rho}+ej^{\mu}=0.
\eea
Hence, the equation shows \cite{Zhang:1988wy}
\bea
\frac{e^2}{2\theta}\epsilon^{0ij}\partial_{i}a_{j}+ej^0=-\frac{e^2}{2\theta}b_i+ej^0=0
\eea
or
\bea
b(\vec{r})=\frac{2\theta}{e}|\psi(\vec{r})|^2
\eea
from the equation of motion of the gauge field $a_0$ and also shows \cite{Zhang:1988wy}:
\bea
\frac{e^2}{2\theta}\epsilon^{i0j}\dot{a}_j+\frac{e^2}{2\theta}\epsilon^{ij0}\partial_ja_0+ej^i&=&\frac{e^2}{2\theta}\epsilon_{ij}\partial_ja_0-\frac{e^2}{2\theta}\epsilon_{ij}\dot{a_j}+ej^i=0,
\nn\\
-\frac{e^2}{2\theta}\epsilon_{ij}\partial_i\dot{a}_j+e\partial_i j^i&=&0
\eea
or 
\bea
\dot{b}_i(\vec{r})=-\frac{2\theta}{e}\partial_i j^i=\frac{2\theta}{e}\frac{\partial}{\partial t}|\psi(\vec{r})|^2
\eea
from the equation of motion of the gauge field $a_i$. Thus, the suitable Lagrangian for the mean-field theory is given by \cite{Zhang:1988wy}
\bea
L_{\mathrm{mf}}&=&\psi^{\dagger}(\vec{r})\bigg(i\partial_0-e\big(A_0(\vec{r})+a_0(\vec{r})\big)\bigg)\psi
-\frac{1}{2m_e}\psi^{\dagger}(\vec{r})\big(-i\nabla-\vec{A}(\vec{r})-e\vec{a}(\vec{r})\big)^2\psi(\vec{r})
\nn\\
&&-\frac{1}{2}\int d^2r^{\prime}\ \frac{e^2}{4\pi\epsilon|\vec{r}-\vec{r}^{\prime}|}\rho(\vec{r})\rho(\vec{r}^{\prime})
\nn\\
&&+\frac{e^2}{4\theta}\epsilon^{\mu\nu\rho}a_{\mu}(\vec{r})\partial_{\nu}a_{\rho}(\vec{r}).
\eea
We can find that the filling factor is $1/(2k+1)$ when the quasi-particle is a boson field \cite{Laughlin:1983fy, Zhang:1988wy} and the filling factor is $1/(2k+2)$ when the quasi-particle is a fermion field \cite{Zhang:1988wy}, where $k$ is a non-negative integer.

\subsection{Globally Defined Quantum Hall Effect}
We demonstrate that the mean-field theory also needs to be careful about the global effect as in the 2+1 dimensional duality web from the case of $\theta=2\pi$ \cite{Seiberg:2016gmd}. We can find that the variation of the dynamical gauge field $a_{\mu}$ gives
\bea
\frac{e^2}{4\pi}\epsilon^{\mu\nu\rho}\partial_{\nu}a_{\rho}
\eea
when the composite fermion field vanishes. Thus, the flux does not obey the Dirac quantization condition. To get the Dirac quantization condition for the flux, we redefine the gauge field $a\equiv 2a^{\prime}$, then the flux of the gauge field $a^{\prime}$ obeys the Dirac quantization condition.

  Let us consider empty Landau level or no electrons, then the Lagrangian of the mean-field should not have any composite fermion. Now we should remain one non-trivial Chern-Simons term
\bea
\frac{e^2}{\theta}\epsilon^{\mu\nu\rho}a^{\prime}_{\mu}(\vec{r})\partial_{\nu}a^{\prime}_{\rho}(\vec{r}),
\eea
but the term should give an incorrect description for the empty vacuum. Thus, we expect that the mean-field theory is only correct locally, but it shows the wrong physics globally. Let us redefine the terms related to the quasi-particle in the Lagrangian
\bea
L_{\mathrm{qp}}\lbrack\psi; a+A\rbrack&\equiv&\psi^{\dagger}(\vec{r})\bigg(i\partial_0-e\big(A_0(\vec{r})+a_0(\vec{r})\big)\bigg)\psi
\nn\\
&&-\frac{1}{2m_e}\psi^{\dagger}(\vec{r})\big(-i\nabla-\vec{A}(\vec{r})-e\vec{a}(\vec{r})\big)^2\psi(\vec{r})
\nn\\
&&-\frac{1}{2}\int d^2r^{\prime}\ \frac{e^2}{4\pi\epsilon|\vec{r}-\vec{r}^{\prime}|}\rho(\vec{r})\rho(\vec{r}^{\prime}).
\eea
Now we write one mean-field theory as
\bea
L_{\mathrm{gmf}}&\equiv&L_{\mathrm{qp}}\big\lbrack\psi; 2a^{\prime}+A\big\rbrack+\frac{e^2}{\theta}\epsilon^{\mu\nu\rho}a^{\prime}_{\mu}(\vec{r})\partial_{\nu}a^{\prime}_{\rho}(\vec{r})
\nn\\
&&-\frac{e^2}{\theta}\int d^3x\ \epsilon^{\mu\nu\rho}b_{\mu}\partial_{\nu}\big(a_{\rho}-2a_{\rho}^{\prime}\big).
\eea
If we integrate out the gauge field $b$, it is equivalent to using 
\bea
a_{\rho}=2a_{\rho}^{\prime}.
\eea
Thus, we can use the relation between two gauge fields in the Lagrangian without affecting the equivalence.

Then we redefine the gauge field as:
\bea
a\equiv\tilde{a}-A, \qquad a^{\prime}\equiv\tilde{a}^{\prime}-b
\eea
and rewrite the action of the mean-field theory:
\bea
S_{\mathrm{gmf}}&\equiv&\int d^3x\ L_{\mathrm{gmf}}
\nn\\
&=&\int d^3x\ L_{\mathrm{qp}}\lbrack\psi; \tilde{a}\rbrack+\frac{e^2}{\theta}\int d^3x\
\epsilon^{\mu\nu\rho}(\tilde{a}^{\prime}_{\mu}-b_{\mu})\partial_{\nu}(\tilde{a}^{\prime}_{\rho}-b_{\rho})
\nn\\
&&-\frac{e^2}{\theta}\int d^3x\ \epsilon^{\mu\nu\rho}b_{\mu}\partial_{\nu}\big(a_{\rho}-2a_{\rho}^{\prime}\big)
\nn\\
&=&\int d^3x\ L_{\mathrm{qp}}\lbrack\psi; \tilde{a}\rbrack+\frac{e^2}{\theta}\int d^3x\ \epsilon^{\mu\nu\rho}\tilde{a}_{\mu}^{\prime}\partial_{\nu}\tilde{a}^{\prime}_{\rho}
\nn\\
&&+\frac{e^2}{\theta}\int d^3x\ \epsilon^{\mu\nu\rho}b_{\mu}\partial_{\nu}b_{\rho}
-\frac{2e^2}{\theta}\int d^3x\ \epsilon^{\mu\nu\rho}b_{\mu}\partial_{\nu}\tilde{a}^{\prime}_{\rho}
\nn\\
&&-\frac{e^2}{\theta}\int d^3x\ \epsilon^{\mu\nu\rho}b_{\mu}\partial_{\nu}\big(a_{\rho}-2a_{\rho}^{\prime}\big)
\nn\\
&=&\int d^3x\ L_{\mathrm{qp}}\lbrack\psi; \tilde{a}\rbrack
\nn\\
&&-\frac{e^2}{\theta}\int d^3x\ \epsilon^{\mu\nu\rho}b_{\mu}\partial_{\nu}b_{\rho}+\frac{2e^2}{\theta}\int d^3x\ \epsilon^{\mu\nu\rho}b_{\mu}\partial_{\nu}b_{\rho}
\nn\\
&&-\frac{2e^2}{\theta}\int d^3x\ \epsilon^{\mu\nu\rho}b_{\mu}\partial_{\nu}\tilde{a}^{\prime}_{\rho}
+\frac{e^2}{\theta}\int d^3x\ \epsilon^{\mu\nu\rho}\tilde{a}_{\mu}^{\prime}\partial_{\nu}\tilde{a}^{\prime}_{\rho}
\nn\\
&&-\frac{e^2}{\theta}\int d^3x\ \epsilon^{\mu\nu\rho}b_{\mu}\partial_{\nu}\big(a_{\rho}-2a_{\rho}^{\prime}\big)
\nn\\
&=&\int d^3x\ L_{\mathrm{qp}}\lbrack\psi; \tilde{a}\rbrack-\frac{e^2}{\theta}\int d^3x\ \epsilon^{\mu\nu\rho}b_{\mu}\partial_{\nu}b_{\rho}
\nn\\
&&-\frac{2e^2}{\theta}\int d^3x\ \epsilon^{\mu\nu\rho}b_{\mu}\partial_{\nu}\big(\tilde{a}^{\prime}_{\rho}-b_{\rho}\big)
-\frac{e^2}{\theta}\int d^3x\ \epsilon^{\mu\nu\rho}b_{\mu}\partial_{\nu}\big(a_{\rho}-2a_{\rho}^{\prime}\big)
\nn\\
&&+\frac{e^2}{\theta}\int d^3x\ \epsilon^{\mu\nu\rho}\tilde{a}_{\mu}^{\prime}\partial_{\nu}\tilde{a}^{\prime}_{\rho}
\nn\\
&=&\int d^3x\ L_{\mathrm{qp}}\lbrack\psi; \tilde{a}\rbrack-\frac{e^2}{\theta}\int d^3x\ \epsilon^{\mu\nu\rho}b_{\mu}\partial_{\nu}b_{\rho}
\nn\\
&&-\frac{e^2}{\theta}\int d^3x\ \epsilon^{\mu\nu\rho}b_{\mu}\partial_{\nu}\big(\tilde{a}_{\rho}-A_{\rho}\big)
+\frac{e^2}{\theta}\int d^3x\ \epsilon^{\mu\nu\rho}\tilde{a}_{\mu}^{\prime}\partial_{\nu}\tilde{a}^{\prime}_{\rho}.
\eea
We can find that action of the mean-field theory has the problematic Chern-Simons term
\bea
\frac{e^2}{\theta}\int d^3x\ \epsilon^{\mu\nu\rho}\tilde{a}_{\mu}^{\prime}\partial_{\nu}\tilde{a}^{\prime}_{\rho},
\eea
but it is decoupled from other terms of the action now. Now we can ignore this term. We can find that if we integrate out the dynamical gauge field $\tilde{a}$ for the empty Landau level, it is equivalent to taking the vanishing gauge field $b=0$ and the theory is trivial. Thus, this theory gives a correct description to the empty Landau level at global sense.

Let us compare differences between two mean-field theories for the empty Landau level. In the mean-field theory $L_{\mathrm{mf}}$, the problematic Chern-Simons term cannot be decoupled so it shows an incorrect description. In the mean-field theory $L_{\mathrm{gmf}}$, we introduce the auxiliary field $b$. This auxiliary field lets the gauge field $\tilde{a}$ not be the same as the gauge field $\tilde{a}^{\prime}$. Hence, the non-trivial Chern-Simons term can be decoupled from the mean-field theory. When we turn off the auxiliary gauge field $b$, the non-trivial Chern-Simons term should exist when we consider the empty Landau level. Hence, this means that the correct description of the empty Landau level relies on that the auxiliary gauge field $b$ lets the problematic Chern-Simons term decouples from the mean-field theories. We can also find that the partition function of these two mean-field theories are also the same without ignoring the decoupled terms for the empty Landau level. Hence, this means that two mean-field theories just give different physical interpretations to the empty Landau level.

\section{Electric-Magnetic Duality in Four Dimensions and the 2+1 Dimensional Duality Web}
\label{6}
We first review the electric-magnetic duality of a four dimensional Abelian gauge theory with boundary \cite{Witten:2003ya}. Then we discuss the 2+1 dimensional duality web \cite{Seiberg:2016gmd} from the electric-magnetic duality \cite{Seiberg:2016gmd} and show that the dualities in 2+1 dimensional duality web is related to the T-transformation and the S-transformation \cite{Seiberg:2016gmd, Witten:2003ya} or a combination of the transformations $\big($SL(2) transformation$\big)$. The perspective of the electric-magnetic duality in the 2+1 dimensional duality web \cite{Seiberg:2016gmd} also shows a consistent study and this study does not rely on the conjecture of the 2+1 dimensional duality web \cite{Seiberg:2016gmd}.

\subsection{Electric-Magnetic Duality in Four Dimensions}
We show the electric-magnetic duality in the 3+1 dimensional Abelian gauge theory and the action of the Abelian gauge theory with the theta term in 3+1 dimensions is 
\bea
S_{1}&=&\int d^{4}x\ \bigg(\frac{1}{4e^2}F_{\mu\nu}F^{\mu\nu}+\frac{\theta}{32\pi^2}\epsilon^{\mu_1\mu_2\nu_1\nu_2}F_{\mu_1\mu_2}F_{\nu_1\nu_2}
\bigg),
\nn\\
\eea
where $F\equiv dA$, $A$ is the one-form potential, and $F$ is the field strength associated to the one-form potential. The action can be written as
\bea
S_1=\frac{i}{8\pi}\int d^{4}x\ \bigg(\tau^*F^+_{\mu_1\mu_2}F^{+,\ \mu_1\mu_2}
-\tau F^-_{\mu_1\mu_2}F^{-,\ \mu_1\mu_2}\bigg),
\eea
where 
\bea
F^{\pm}_{\mu_1\mu_2}&\equiv&\frac{1}{2}\bigg(F_{\mu_1\mu_2}\pm\frac{i}{2}\epsilon_{\mu_1\mu_2\nu_1\nu_2}F^{\nu_1\nu_2}\bigg), \qquad
\tau\equiv-\frac{\theta}{2\pi}+\frac{2\pi i}{e^2}.
\eea
The first term of  the action $S_1$ is written as:
\bea
&&\tau^*F^+_{\mu_1\mu_2}F^{+,\ \mu_1\mu_2}
\nn\\
&=&\bigg(-\frac{\theta}{2\pi}-\frac{2\pi i}{e^2}\bigg)\frac{1}{4}\bigg(F_{\mu_1\mu_2}+\frac{i}{2}\epsilon_{\mu_1\mu_2\nu_1\nu_2}F^{\nu_1\nu_2}\bigg)
 \bigg(F^{\mu_1\mu_2}+\frac{i}{(p+1)!}\epsilon^{\mu_1\mu_2\nu_1\nu_2}F_{\nu_1\nu_2}\bigg)
\nn\\
&=&\frac{1}{4}\bigg(-\frac{\theta}{2\pi}-\frac{2\pi i}{e^2}\bigg)\bigg(F_{\mu_1\mu_2}F^{\mu_1\mu_2}
+F_{\mu_1\mu_2}F^{\mu_1\mu_2}+i\epsilon_{\mu_1\mu_2\nu_1\nu_2}
F^{\mu_1\mu_2}F^{\nu_1\nu_2}\bigg)
\nn\\
&=&\frac{1}{2}\bigg(-\frac{\theta}{2\pi}-\frac{2\pi i}{e^2}\bigg)F_{\mu_1\mu_2}F^{\mu_1\mu_2}
+\frac{1}{4}\bigg(\frac{2\pi}{e^2}-\frac{i\theta}{2\pi}\bigg)\epsilon_{\mu_1\mu_2\nu_1\nu_2}
F^{\mu_1\mu_2}F^{\nu_1\nu_2}.
\eea
Thus, we obtain:
\bea
&&\tau^*F^+_{\mu_1\mu_2}F^{+,\ \mu_1\mu_2}-\tau F^-_{\mu_1\mu_2}F^{-,\ \mu_1\mu_2}
=-\frac{\pi i}{e^2}F_{\mu_1\mu_2}F^{\mu_1\mu_2}
-\frac{i\theta}{4\pi}\epsilon_{\mu_1\mu_2\nu_1\nu_2}
F^{\mu_1\mu_2}F^{\nu_1\nu_2},
\nn\\
\eea
\bea
S_1&=&\frac{i}{8\pi}\int d^{4}x\ \bigg(\tau^*F^+_{\mu_1\mu_2}F^{+,\ \mu_1\mu_2}
-\tau F^-_{\mu_1\mu_2}F^{-,\ \mu_1\mu_2}\bigg)
\nn\\
&=&\int d^{4}x\ \bigg(\frac{1}{4e^2}F_{\mu_1\mu_2}F^{\mu_1\mu_2}
+\frac{\theta}{32\pi^2}\epsilon^{\mu_1\mu_2\nu_1\nu_2}
F_{\mu_1\mu_2}F_{\nu_1\nu_2}
\bigg).
\eea

Now we perform the electric-magnetic duality by introducing the auxiliary field $G$ to the action
\bea
S_1&\rightarrow&\int d^{4}x\ \bigg\lbrack\frac{1}{4e^2}F_{\mu_1\mu_2}F^{\mu_1\mu_2}
+\frac{\theta}{32\pi^2}\epsilon^{\mu_1\mu_2\nu_1\nu_2}
F_{\mu_1\mu_2}F_{\nu_1\nu_2}
\nn\\
&&+\frac{1}{8\pi}\epsilon^{\mu_1\mu_2\nu_1\nu_2}G_{\mu_1\mu_2}
\bigg(F_{\nu_1\nu_2}-\big(\partial_{\nu_1}A_{\nu_2}-
\partial_{\nu_2}A_{\nu_1}\big)\bigg)\bigg\rbrack.
\eea
We can integrate out the two-form $F$ and it is equivalent to using
\bea
&&\frac{1}{2e^2}F_{\mu_1\mu_2}+\frac{\theta}{16\pi^2}\epsilon_{\mu_1\mu_2\nu_1\nu_2}F^{\nu_1\nu_2}
+\frac{1}{8\pi}\epsilon_{\mu_1\mu_2\nu_1\nu_2}G^{\nu_1\nu_2}=0
\eea
or
\bea
-\frac{\theta}{4\pi^2}F_{\mu_1\mu_2\cdots\mu_{p+1}}+\frac{1}{2e^2}\epsilon_{\mu_1\mu_2\nu_1\nu_2}F^{\nu_1\nu_2}-\frac{1}{2\pi}G_{\mu_1\mu_2}=0.
\eea
In other words, we can equivalently use
\bea
&&\bigg(\frac{1}{2e^4}+\frac{\theta^2}{32\pi^4}\bigg)F_{\mu_1\mu_2}+\bigg(\frac{\theta}{16\pi^3}G_{\mu_1\mu_2}
+\frac{1}{8e^2\pi}\epsilon_{\nu_1\mu_2\nu_1\nu_2}G^{\nu_1\nu_2}\bigg)=0
\eea
or
\bea
F_{\mu_1\mu_2}&=&-\frac{\pi}{e^2\bigg(\frac{4\pi^2}{e^4}+\frac{\theta^2}{4\pi^2}\bigg)}\epsilon_{\mu_1\mu_2\nu_1\nu_2}G^{\nu_1\nu_2}
-\frac{\frac{\theta}{2\pi}}{\frac{\theta^2}{4\pi^2}+\frac{4\pi^2}{e^4}}G_{\mu_1\mu_2}.
\eea
After we perform the electric-magnetic duality, we obtain the action
\bea
&&\int d^{4}x\ \bigg(\frac{1}{4e^2}G_{\mu_1\mu_2}G^{\mu_1\mu_2}-\frac{\theta}{32\pi^2}\epsilon^{\mu_1\mu_2\nu_1\nu_2}G_{\mu_1\mu_2}
G_{\nu_1\nu_2}\bigg)\frac{1}{\frac{\theta^2}{4\pi^2}+\frac{4\pi^2}{e^4}}.
\eea
Thus, the electric-magnetic duality transformation shows:
\bea
\tau\rightarrow\frac{\frac{\theta}{2\pi}}{\frac{\theta^2}{4\pi^2}+\frac{4\pi^2}{e^4}}+\frac{\frac{2\pi i}{e^2}}{\frac{\theta^2}{4\pi^2}+\frac{4\pi^2}{e^4}}=\frac{1}{\frac{\theta}{2\pi}-\frac{2\pi i}{e^2}}=-\frac{1}{\tau}.
\eea
We can find two transformations, in which a combination of the transformations is the SL(2) transformation. The first transformation is the S-transformation
\bea
S(\tau)=-\frac{1}{\tau}
\eea
and the second transformation is the T-transformation
\bea
T(\tau)=\tau+1,
\eea
which generates the transformation
\bea
\theta\rightarrow\theta+2\pi.
\eea

Now we consider a manifold with a boundary manifold $M$. The action is 
\bea
S_{\mathrm{1b}}&=&\frac{1}{2\pi}\int_Md^{3}x\sqrt{\det{g_{\mu\nu}}}\ J_{\mu}B^{\mu}
+\frac{i}{8\pi}\int d^{4}x\ \bigg(\tau^*F^+_{\mu_1\mu_2}F^{+,\ \mu_1\mu_2}
-\tau F^-_{\mu_1\mu_2}F^{-,\ \mu_1\mu_2}\bigg),
\nn\\
\eea
where $B$ is a one-form gauge field lives on the boundary manifold $M$ and is also a boundary value of the one-form gauge field $A$, which is a gauge potential associated to the field strength $F$ and $J$ is a conserved current on the three dimensional manifold $M$. Because an equation of the motion of the gauge potential $B$ on the boundary $M$ depends on the gauge potential $B$ rather than a field strength, we only perform the electric-magnetic duality in the bulk space with a fixed boundary value of the gauge field $A$, which is not changed by the electric-magnetic duality \cite{Witten:2003ya}.

Thus, we add the term 
\bea
&&\frac{1}{8\pi}\int d^{4}x\
\epsilon^{\mu_1\mu_2\nu_1\nu_2}G_{\mu_1\mu_2}\bigg(F_{\nu_1\nu_2}-\big(\partial_{\nu_1}A_{\nu_2}-
\partial_{\nu_2}A_{\nu_1}\big)\bigg)
\eea
to the action $S_{\mathrm{1b}}$
and the field strength $F$ becomes an arbitrary two-form without any restriction and $G$ is also an arbitrary two-form. We first integrate out the two-form field $F$ in the bulk region and obtain
\bea
&&\frac{i}{8\pi}\int d^{4}x\ \bigg(\tau^{\prime *}G^+_{\mu_1\mu_2}G^{+,\ \mu_1\mu_2}
-\tau^{\prime} G^-_{\mu_1\mu_2}G^{-,\ \mu_1\mu_2}\bigg)
\nn\\
&&-\frac{1}{8\pi}\int d^{4}x\
\epsilon^{\mu_1\mu_2\nu_1\nu_2}G_{\mu_1\mu_2}
\big(\partial_{\nu_1}A_{\nu_2}-
\partial_{\nu_2}A_{\nu_1}\big)
\nn\\
&&+\frac{1}{2\pi}\int_Md^{3}x\sqrt{\det{g_{\mu\nu}}}\ J_{\mu_1}B^{\mu_1},
\eea
where
\bea 
\tau^{\prime}\equiv-\frac{1}{\tau}.
\eea 
Now $G$ is not an arbitrary two-form after we perform the electric-magnetic duality. Thus, we obtain the BF term
\bea
&&-\frac{1}{8\pi}\int d^{4}x\
\epsilon^{\mu_1\mu_2\nu_1\nu_2}G_{\mu_1\mu_2}
\big(\partial_{\nu_1}A_{\nu_2}-
\partial_{\nu_2}A_{\nu_1}\big)
\nn\\
&\rightarrow&-\frac{1}{2\pi}\int_M d^{3}x\ \epsilon^{\mu_1\mu_2\mu_{3}} B_{\mu_1}\partial_{\mu_{2}}A_{\mu_{3}}.
\eea
In other words, the S-transformation shows:
\bea
\tau\rightarrow -\frac{1}{\tau}, \qquad F\rightarrow G,
\eea
and provides the BF term on the boundary manifold $M$.

The T-transformation ($\theta\rightarrow\theta+2\pi$) gives the Chern-Simons theory
\bea
\frac{1}{4\pi}\int_Md^{3}x\ \epsilon^{\mu_1\mu_2\mu_{3}}B_{\mu_1}\partial_{\mu_{2}}
B_{\mu_{3}}
\eea
from the theta term. This shows that the T-transformation is not a symmetry in the bulk so we need to consider a combination of the bulk theory and the boundary theory. Thus, the boundary theory under the T-transformation should give the Chern-Simons theory
\bea
-\frac{1}{4\pi}\int_Md^{3}x\ \epsilon^{\mu_1\mu_2\mu_{3}}B_{\mu_1}\partial_{\mu_{2}}
B_{\mu_{3}}.
\eea
to show invariance under the T-transformation. The study can also be extended to the $p$-form gauge field in $2p+2$ dimensions with a boundary manifold \cite{Ma:2017mpb}.

\subsection{Conjecture of the 2+1 Dimensional Duality Web and the 3+1 Dimensional Electric-Magnetic Duality}
To interpret the conjecture of the 2+1 dimensional duality web from the 3+1 dimensional electric-magnetic duality, we consider two theories. The action of the first theory is
\bea
S_{\mathrm{I}}=S_1\lbrack A; \tau\rbrack+\int_M d^3x\sqrt{\det{g_{\mu\nu}}}\ i\bar{\psi}\gamma^{\mu_1}(\partial_{\mu_1}+iA_{\mu_1})\psi
\eea
where 
\bea
\tau\equiv -\frac{\theta}{2\pi}+\frac{2\pi i}{e^2}.
\eea
The action of the second theory is
\bea
S_{\mathrm{II}}=S_1\lbrack B; \tau^{\prime}\rbrack+\int_Md^3x\sqrt{\det{g_{\mu\nu}}}\ \big(|(\partial_{\mu_1}+iB_{\mu_1})\phi|^2-\lambda|\phi|^4\big),
\eea
where the gauge field $B$ is the U(1) gauge field, and $\tau^{\prime}$ is defined by 
\bea
\tau^{\prime}\equiv 1-\frac{1}{\tau}.
\eea
If we consider the weak coupling limit $e^2\rightarrow 0$, the gauge field $A$ in the first theory becomes the background field, but the gauge field $B$ in the second theory is still dynamical. Hence, the first theory reduces to the 2+1 dimensional Dirac fermion theory with a spin$_c$ background gague field $A$. Now we assume that these two theories are the same and show that the SL(2) transformation can lead the assumption to the conjecture of the 2+1 dimensional duality web by taking the weak coupling limit $e^2\rightarrow 0$. 

We first perform the $\mathrm{T}^{-1}$-transformation on the second theory, then the action of the second theory becomes
\bea
&&S_1\bigg\lbrack B;-\frac{1}{\tau}\bigg\rbrack+\int_Md^3x\sqrt{\det{g_{\mu\nu}}}\ \big(|(\partial_{\mu_1}+iB_{\mu_1})\phi|^2-\lambda|\phi|^4\big)
\nn\\
&&+\frac{1}{4\pi}\int_M d^3x\ \epsilon^{\mu_1\mu_2\mu_3}B_{\mu_1}\partial_{\mu_2}B_{\mu_3}.
\eea
Then we perform the $\mathrm{S}^{-1}$-transformation and the action becomes
\bea
&&S_1\lbrack A^{\prime}; \tau\rbrack+\int_Md^3x\sqrt{\det{g_{\mu\nu}}}\ \big(|(\partial_{\mu_1}+iB_{\mu_1})\phi|^2-\lambda|\phi|^4\big)
\nn\\
&&+\frac{1}{4\pi}\int_M d^3x\ \epsilon^{\mu_1\mu_2\mu_3}B_{\mu_1}\partial_{\mu_2}B_{\mu_3}
+\frac{1}{2\pi}\int_M d^3x\ \epsilon^{\mu_1\mu_2\mu_3}B_{\mu_1}\partial_{\mu_2}A^{\prime}_{\mu_3},
\eea
where $A^{\prime}$  is the spin$_c$ gauge field. Because the gauge field $A$ in the first theory is also the spin$_c$ gauge field, we can identify these two gauge fields $A=A^{\prime}$. Now $\tau^{\prime}$ becomes $\tau$ so taking the weak coupling limit lets the gauge field $A$ become the background gauge field and the theory also reduces to the boson theory of the 2+1 dimensional duality web. Hence, we obtain the conjecture of the 2+1 dimensional duality web 
\bea
i\bar{\psi}\slashed{D}_A\psi\longleftrightarrow |D_a\phi|^2-\lambda|\phi|^4+\frac{1}{4\pi}ada+\frac{1}{2\pi}adA
\eea
from the four dimensional electric-magnetic duality. Other dualities of the 2+1 dimensional duality web also has corresponding four dimensional dualities \cite{Seiberg:2016gmd}. Hence, this study uses the four dimensional electric-magnetic duality reinterprets the 2+1 dimensional duality web. From the four dimensional perspective, it is unnecessary to use the conjecture of the 2+1 dimensional duality web. Hence, this consistent study gives other evidences to the conjecture of the 2+1 dimensional duality web.  

Because the 2+1 dimensional duality web should be correct at the IR limit, we should expect that the gauge coupling constant should be weak in the four dimensional theories from the renormalization group flow. Hence, this also interprets that the three dimensional background gauge fields comes from the four dimensional gauge fields if the four dimensional interpretation is correct.

\section{Outlook and Future Directions}
\label{7}
The Dirac fermion theory has parity anomaly in odd dimensions so it is a non-gauge invariant theory at quantum level \cite{Niemi:1983rq}. Hence, this theory is not gauge invariant so the parity anomaly needs to be canceled from a combination of the Dirac fermion theory and one theory. Based on the argument of the charges, one can build the conjecture of the 2+1 dimensional duality web \cite{Karch:2016sxi} at the IR limit. The conjecture of the 2+1 dimensional duality web lead the particle-vortex duality of bosons \cite{Peskin:1977kp} and fermions \cite{Metlitski:2015eka} at the IR limit and the particle-vortex dualities can be shown without the conjecture of the 2+1 dimensional duality web at the IR limit. The conjecture leads other dualities to form the interesting 2+1 dimensional duality web at the IR limit. The conjecture of the 2+1 dimensional duality web at the IR limit can also be consistently understood from the perspective of the four dimensional electric-magnetic duality. Thus, the conjecture of the 2+1 dimensional duality web possibly be realizable at the IR limit. The above stories only show the duality web at the zero temperature limit and in the 2+1 dimensions. Thus, the other generalizations are still an open question in the duality web at the IR limit.

When we study non-zero temperature or go beyond 2+1 dimensions, the first issue of these generalizations should be the spin structure. However, the duality web may not be guaranteed for all backgrounds. If we restrict our study to the flat background, the generalizations are still possible. 

For the non-zero temperature in 2+1 dimensions, the Dirac fermion theory generates the Chern-Simons term with a temperature dependent coefficient at one-loop. If we directly apply the same way to the 2+1 dimensional duality web, we should find a non-gauge invariant issue because the coefficient depends on a temperature in the low-energy effective theory or the Chern-Simons term \cite{Deser:1997nv, Ma:2016yas, Ma:2017mpb}. However, the non-gauge invariant term appears because we only work the perturbation at the leading order \cite{Deser:1997nv}. If we know how to do non-perturbative computation in the Dirac fermion theory as in the one dimensional Dirac fermion theory, the non-gauge invariant issue should go away \cite{Fosco:1997vu}. Hence, we expect that dualities should exist at a non-vanishing temperature at least for the leading order with respect to a number of gauge field. To show more consistencies at a non-zero temperature, we need to go beyond the low-energy limit and check the dualities. This is still not done so far. The low-energy physics should be interesting for the phase transition. Thus, a generalization of the duality web to a non-zero temperature should let us study phase transitions with respect to a temperature.

Going beyond the 2+1 dimensional duality web at zero temperature, a simple way is to try electric-magnetic duality in a non-interacting $p$-form theory in $2p+2$ dimensions \cite{Ma:2017mpb}. Then this should indicate what terms should be coupled to fermion theories or boson theories if the electric-magnetic duality is still useful for building an odd dimensional duality web. However, the direction is still unclear. If a higher dimensional duality web can be constructed, these theories should have corresponding low dimensional physics through compactification. Hence, it is interesting to find more duality webs in three dimensions and four dimensions from this way. 

The duality web is only constructed at the IR limit. Hence, this leaves many ambiguities there. For example. we can also consider the four dimensional non-interacting massless Dirac fermion theory with a boundary condition. Then the boundary condition can generate the parity anomaly in a three dimensional boundary manifold \cite{Kurkov:2018pjw}. Thus, it is a different way to rewrite the fermion theory of the conjecture of the 2+1 dimensional duality web at the IR limit. Hence, we should find a way to construct a duality web beyond the IR limit to avoid such ambiguity.   

The physical system in the three dimensional spin$_c$ manifold also provides the global effect, which does not present in the three dimensional spin manifold. This global effect is overlooked without the perspective of the 2+1 dimensional duality web. As we demonstrated, the global effect is also important in the practical system, 2+1 dimensional quantum Hall effect. This should show that the global effect in the 2+1 dimensional duality web should be very practical and the effect can be observed because these studies at the IR limit as in the 2+1 dimensional quantum Hall effect. This also gives us a new theoretical perspective to topological quantum theory and condensed matter theories.

\section*{Acknowledgments}
The author would like to thank Chang-Tse Hsieh and Ken Shiozaki for their useful discussion and Nan-Peng Ma for his encouragement.

\end{document}